\documentclass[a4paper]{JHEP3}
\usepackage{amssymb}
\usepackage{amsmath}
\usepackage{cite}
\usepackage{graphicx}

\def\be{\begin{equation}}
\def\ee{\end{equation}}
\def\baray{\begin{eqnarray}}
\def\earay{\end{eqnarray}}
\def\ba{\begin{eqnarray}}
\def\ea{\end{eqnarray}}

\def\pdfannotlink{\pdfstartlink}

\title{Bispectrum signatures of a modified vacuum
in single field inflation with a small speed of sound}

\author{P. Daniel Meerburg$^{1}$, Jan Pieter
van der Schaar$^{2}$ and Mark G. Jackson$^3$\\
$^1$ Astronomical Institute ``Anton Pannekoek", University of Amsterdam,\\
Science Park 904, 1098XH Amsterdam, The Netherlands\\
$^2$  Korteweg-de Vries Institute for Mathematics, University of Amsterdam,\\ 
Plantage Muidergracht 24, 1018 TV Amsterdam, The Netherlands \\
$^3$ Lorentz Institute for Theoretical Physics, 2333 CA Leiden, The Netherlands\\}
\date{\today}

\abstract{Deviations from the Bunch-Davies vacuum during an inflationary period can leave a testable imprint on the higher-order correlations of the CMB and large scale structures in the Universe.  The effect is particularly pronounced if the statistical non-Gaussianity is inherently large, such as in models of inflation with a small speed of sound, e.g. DBI.  First reviewing the motivations for a modified vacuum, we calculate the non-Gaussianity for a general action with a small speed of sound.  The shape of its bispectrum is found to most resemble the `orthogonal' or `local' templates depending on the phase of the Bogolyubov parameter. In particular, for DBI models of inflation the bispectrum can have a profound `local' template feature, in contrast to previous results. Determining the projection into the observational templates allows us to derive constraints on the 
absolute value of the Bogolyubov parameter. In the small sound speed limit, the derived constraints are generally stronger than 
the constraint obtainable from the power spectrum. The bound on the absolute value of the Bogolyubov parameter ranges from
 the $10^{-6}$ to the $10^{-3}$ level for $H/\Lambda_c=10^{-3}$, depending on the specific details of the model, the sound speed and the phase of 
 the Bogolyubov parameter.}

\keywords{Inflation, Cosmic Microwave Background, non-Gaussianities, Dirac-Born-Infeld, Bunch-Davies}

\begin{document}
\section{Introduction}
Recent years have witnessed a steady increase in the amount of precision data from the cosmic microwave background and large scale structure \cite{WMAP5}.  This opportunity has motivated the study of both the power spectrum (or two-point correlation) of the primordial density perturbations, essentially probing the free theory of the inflationary Lagrangian, as well as higher-order statistics, probing the interacting theory. The leading statistical probe of interactions is the bispectrum (or three-point correlation, indicating non-Gaussian statistics) and there is currently a large collection of work on the non-Gaussian predictions for specific models of inflation \cite{ng-singlefield, ng-multifield, ng-curvaton, ng-ekpyrosis, ng-misc}. Interestingly, any detection of non-Gaussianity will immediately rule out all models of single-field slow-roll inflation.  While this would be a monumental discovery in itself, it would also be the starting point for further constraining the inflationary Lagrangian.  Indeed, if the non-Gaussian amplitude is high enough to be observable it opens up a new, and very distinctive, window on the details of inflation \cite{Komatsu:2009kd}.

In a previous article by two of us (PDM, JPvdS) \cite{Meerburg} we computed the primordial bispectrum for a single-field slow-roll inflationary model, with the assumption that the initial state is not the Bunch-Davies (BD) vacuum \cite{Bunch:1978yq}.  This assumption is well-motivated because inflation is by construction an effective theory and therefore there is no `knowledge' of anything beyond a specified cutoff scale in energy.  In expanding backgrounds this energy-cutoff is tantamount to a (possibly momentum-dependent) time-cutoff, allowing one to instead parameterize ignorance as an initial state from which to evolve the dynamics.  As with all effective theories, the details underlying the construction of such a state are irrelevant; the only concern is to use this state to compute the consequences on the bispectrum, and if possible compare theoretical predictions with the data to place constraints on such a deviation.  

As first reported in \cite{Holman2007},  our results in \cite{Meerburg} confirmed that small initial state modifications can induce surprisingly large corrections to the primordial bispectrum, especially if one includes higher-derivative operators (see also \cite{ColHol}).  The intuitive explanation is essentially that any deviation from BD introduces particles on sub-horizon scales at the initial stages of inflation, and these particles then have ample time to produce a large non-Gaussian signal in the presence of (irrelevant) interactions.  Unfortunately, this yielded only a rather weak bound of order $|\beta|<10^{-2}$ on the deviations from the BD state in terms of the Bogolyubov $\beta$-parameter. This is because although the \emph{magnitude} of such induced non-Gaussianity was large, its \emph{shape} in co-moving momentum space was found to substantially depart from standardized observational templates, providing poor overlap with existing constraints. This motivates the need for better templates and possible systematic analysis of oscillatory features in the bispectrum, for which we hope to report some progress in the future. 

In this article we generalize the class of models for which initial state modifications are considered.  We opine that initial state modifications are generally natural for these models.  In particular, we will argue that for both branches in DBI inflation, UV \cite{AST:UVDBI} and IR \cite{Chen:IRDBI,Chen:2005fe}, small deviations can be expected or at least are worth considering when computing the full bispectrum.  Our starting point will be the most general single-field action, for which the bispectrum has been previously computed in \cite{SeeLid, Chenetal}.  Specific single-field models (slow-roll, DBI, kinetic) are all limiting cases of this action, and we will be interested in the limit where the speed of sound $c_s$ (that is, the speed at which inflaton perturbations evolve) is significantly reduced relative to $c$, the speed at which metric perturbations evolve. Whereas bispectrum components in standard slow-roll models are typically proportional to $\epsilon$ and $\eta$ \cite{Maldacena}, in the limit of small $c_s$ one finds new contributions to the bispectrum that are proportional to $1/c_s^2$.  Assuming the BD-vacuum these potentially large corrections are typically of the equilateral type and have been previously constrained \cite{CSZT2007, SSZ2009}.  When one allows for (initially Gaussian) vacuum state modifications, there are yet-additional components to the bispectrum which are also expected to be enhanced in the small speed of sound limit. As a consequence, we will be able to put relatively strong constraints on $|\beta|$. Importantly, we will also study the effect of an arbitrary phase in the Bogolyubov parameter, which has so far been neglected.  Somewhat surprisingly, we will find that specific choices for the phase can have dramatic effects on the shape of the leading non-Gaussian contribution. 
As a consequence the phase of the complex Bogolyubov parameter $\beta$ strongly affects the observational constraints on the absolute value $|\beta|$ 
that can be derived. An important restriction in our analysis is that we will have to assume approximate scale-invariance throughout and will not 
discuss (big or small) departures from scale invariance and how these can be observationally constrained \cite{Khoury, NG-scaledep}. 

This article is organized as follows.  In \S 2 we will briefly summarize the small-$c_s$ models of inflation and explain how initial state modifications are quite natural in this context.  In \S 3 we begin from the most general single-field inflationary action and then compute the corrections to the bispectrum.  In \S 4 we identify the leading-order and sub-leading components and then constrain modifications to the initial state.  Constraints can only be set using already measured bispectra by comparing the theoretical components to observed templates.  In order to achieve this, we compute the leakage via projection of these components into measured components, for which we make use of a formalism first proposed in \cite{BCZ2004} and optimized in \cite{Fergusson}.  In \S 5 we use the two derived constraints on different $f_{\mathrm{NL}}$ to derive constraints on DBI models with three free parameters.  In \S 6 we summarize and conclude.

\section{A case for vacuum state modifications in small sound speed models}

It has now become clear that vacuum state modifications cannot be excluded on purely theoretical grounds, even though they are constrained by bounds on backreaction \cite{GKSS-br, Mottola-br} and observations of the primordial power spectrum \cite{EKP1, GKSS, EKP2, WMAP3yr, Slothetal}. The theoretical challenge is to relate potential vacuum state modifications to ultraviolet physics at the string- or Planck-scale. In terms of an effective field theory description of inflation such a relation would appear natural, since the details of the vacuum (or equivalently initial) state of inflation usually depends on physics far beyond the cutoff in the effective field theory description. This raises the hope of being able to probe string- or Planck-scale physics through the signatures of possible modifications to the vacuum state. 

Unfortunately, so far the available proposals have been phenomenological in nature and can be separated into two classes. In the original work of Danielsson \cite{NPH} scale-invariance is preserved by proposing that the initial time $\eta_0$ (beyond which an effective field theory theory breaks down) is identified as the time where the physical momentum equals some high energy cutoff, defining a New Physics Hypersurface (NPH). This immediately implies that the initial ``cutoff" time will be different for different comoving momenta, i.e. $\eta_0$ will be a function of $k$ allowing for the preservation of scale-invariance. A specific, but rather ad-hoc, prediction for the Bogolyubov parameter is then obtained by assuming a local empty state at the initial time $\eta_0(k)$ \cite {NPH, GKES}.  

In a different approach known as Boundary Effective Field Theory (BEFT) \cite{BEFT}, scale-invariance is explicitly broken by identifying an initial time surface $\eta_0$ where the usual BD boundary conditions are corrected by including leading corrections. As one would expect from effective field theory, the corrections to the BD state grow as a function of comoving momentum $k$, with the largest comoving momenta corresponding to physical momenta closest to the cutoff scale on the initial time surface. The BEFT approach has the advantage of allowing one to calculate corrections to the BD state systematically from first principles. However, the breaking of scale-invariance and the dependence upon the initial time $\eta_0$ are serious drawbacks.

In both scenarios the magnitude of the (leading) corrections, parameterized by the absolute value of the Bogolyubov parameter, is proportional to the inflationary Hubble parameter $H$ divided by a physical cutoff scale $\Lambda_c$. The appearance of the Hubble scale is natural for various reasons and can in particular be related to the fact that an initial vacuum state can only be defined in the adiabatic regime, i.e. for physical momenta much bigger than the Hubble scale.  This immediately suggests that any modification should be proportional to $H / \Lambda_c$, ensuring that the breakdown of a small (perturbative) initial state correction coincides with the breakdown of adiabatic behavior. 

This brings us to our first general observation regarding inflationary models with a small speed of sound. As a direct consequence of the small speed of sound the adiabatic regime breaks down long before the Hubble scale is reached. Instead, the relevant scale in this case is $H^{\star} \equiv H/c_s \gg H$, whose inverse corresponds to the radius of the sound horizon. Consequently, a vacuum state modification is expected to be proportional to $H^{\star}$ in any small speed of sound scenario. Therefore irrespective of a particular motivation or method to describe vacuum state modifications in models with a small speed of sound, its magnitude is anticipated to be enhanced by a factor of $1/c_s$ as compared to standard slow-roll inflationary models (for the same Hubble parameter). Taking into account the constraint on $c_s$ from bispectrum data, which roughly corresponds to $c_s \geq 10^{-2}$ \cite{SSZ2009}, this could add two orders of magnitude to the observable range of $H/ \Lambda_c$ as compared to standard slow-roll inflation. 

The identification of the sound horizon as the relevant scale, signaling the breakdown of the adiabatic approximation, also has a more fundamental consequence for the initial vacuum state.  Generically, the speed of sound slowly evolves during inflation and the models can be separated in two classes: ones where the speed of sound decreases (as in UV DBI models) or models where the speed of sound increases (as in IR DBI models)  towards the end of inflation. In the second case, as was first pointed out in \cite{KinTzi}, when $\epsilon > 1-s$ (where $\epsilon$ is the usual slow-roll parameter and 
$s \equiv \dot{c}_s/H\, c_s$ measures the rate of change in the speed of sound) the fluctuation modes evolve from superhorizon to subhorizon, which excludes the possibility of an adiabatic initial state such as the Bunch-Davies state. Of course, this also rules out any phenomenological approaches to vacuum state modifications that require an (approximately) adiabatic regime, such as the NPH and BEFT proposals. Instead the (different) initial vacuum state in this case should depend crucially on the physics before the inflationary phase. In IR models of DBI inflation this would correspond to the region near the tip of the throat where (strongly coupled) stringy physics is expected to become important. 

\begin{figure}
 \centering
 \includegraphics[scale=0.55]{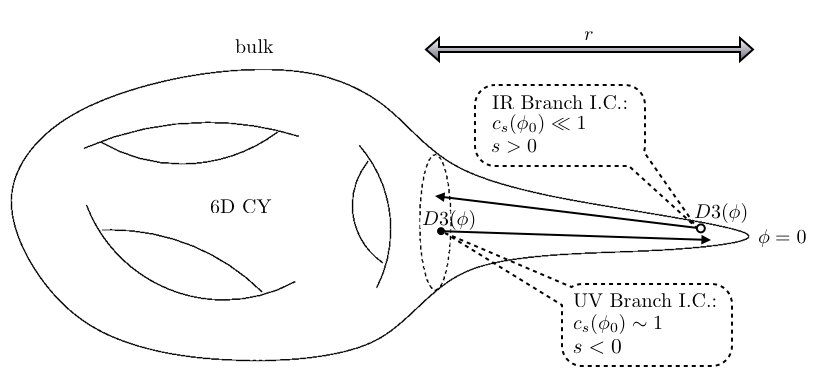} 
 \caption{DBI inflation; IR and UV branches resulting in different evolution for $c_s$, leading to different motivations for changes in the Initial Conditions (I.C.)}
 \label{fig:DBI}
\end{figure}

In the opposite scenario, where the sound speed is decreasing towards the end of inflation, there is also an additional reason why the vacuum could be different from the standard BD vacuum \cite{KinTzi}. In these models the speed of sound at the time of horizon crossing of the scales observable today should be large enough (roughly $c_s >10^{-2}$) to avoid violating current bounds on non-Gaussianity. Depending on the values of the ``slow-roll" parameter $\epsilon$ and the sound speed rate of change $s$, this requirement can imply that there are not enough e-folds left for the largest observable scales to allow for an approximately adiabatic vacuum choice. In other words, some modes in this case are not in the adiabatic regime when inflation began and are probing, just as in the previous case, the phase before inflation started. In the specific example of UV models of DBI inflation (whose simplest versions are in fact already ruled out), these modes would be probing the bulk region outside of the throat, which is governed by stringy physics (see figure \ref{fig:DBI}).

Finally, considering in particular DBI models of inflation, it is important to realize that the DBI action is obtained as the effective theory of open strings after heavy closed strings have been integrated out. As was pointed out in \cite{BurCliHol} integrating out heavy (closed string) states during an inflationary stage could give rise to modified initial conditions in the effective theory of the light (open string) degrees of freedom.

The arguments above provide a relatively strong case for considering departures from the standard BD vacuum in models with a small speed of sound, in addition to the generic effective field theory motivation. It should therefore be worthwhile to look for bispectrum signatures of modified initial states in the limit of a small sound speed, especially as compared to canonical slow-roll inflation. 
In our analysis we will consider initial vacuum state modifications parameterized by an arbitrary complex Bogolyubov parameter, so no particular
model for vacuum state modifications is assumed.  
As first reported in \cite{Holman2007}, and corroborated in \cite{Meerburg, ColHol}, the bispectrum is significantly enhanced due to the 
presence of particles in the modified initial vacuum state. The important requirement of scale-invariance in order to do our analysis, in addition 
to the need for a (comoving dependent) cutoff scale to regulate the integrals after a change in the initial vacuum state, implies that the enhancement 
factor is expected to be proportional to a power of the ratio $\frac{\Lambda}{H^\star}$. This is explains why the bispectrum can give rise to 
relatively strong constraints as compared to the power spectrum. Even though the appearance of $H^\star \equiv H/c_s$ implies that the enhancement is reduced (for fixed $H$) in the small sound speed limit, the specific powers of the enhancement factor that will appear, together with the expected overall enhancement of the non-Gaussian signal in the small sound speed limit, will result in a bispectrum whose 
magnitude is significantly larger as compared to the $c_s=1$ case.

\section{Corrections to the bispectrum}
\subsection{General single-field models of inflation}

We will be working in the context of a general (minimally coupled) 4d action \begin{equation}
S = \frac{1}{2}\int d^{4}x\sqrt{-g}[M_{p}^{2}R-2P(X,\phi)],\label{eq:action}\end{equation}
where $\phi$ is the inflaton field and $X=g^{\mu\nu}\partial_{\mu}\phi\partial_{\nu}\phi$.  From this action we can obtain expressions for the usual slow-roll parameters, as well as a few parameters relevant for the three-point function first introduced in \cite{Chenetal, SeeLid} which
are proportional to higher-order derivatives of $P(X,\phi)$:
\begin{eqnarray}
\epsilon & = & \frac{XP_{,X}}{M_p^2 H^{2}}, \nonumber \\
c_{s}^{2} & = & \frac{P_{,X}}{P_{,X}+2XP_{,XX}} = \frac{M_p^2 H^2 \epsilon}{\Sigma}, \nonumber \\
\Sigma & = & XP_{,X}+2X^{2}P_{,XX}, \nonumber \\
\lambda & = & X^{2}P_{,XX}+\frac{2}{3}X^{3}P_{,XXX} = \frac{1}{3} \left( X \frac{\partial \Sigma}{\partial X} - \Sigma \right) .
\end{eqnarray}
In a recent paper by Senatore \emph{et al.} \cite{SSZ2009} the description of single-field inflationary fluctuations was generalized further by focusing on the (broken) time-translation symmetry and writing down the most general effective Lagrangian of the corresponding Goldstone boson degree of freedom. In their description contributions to the three-point function are organized differently and known results derived in the standard formalism are only reproduced after appropriately translating the relevant variables. In particular they introduce a parameter $\tilde{c}_3$ relevant for the (shape of the) three-point function, which is related to the above set of parameters as
\begin{equation}
\tilde{c}_{3}=\frac{3}{2}c_{s}^{2}\left[\frac{2\lambda}{\Sigma(1-c_{s}^{2})}-1\right].
\label{GBP}
\end{equation}
As pointed out in \cite{SSZ2009}, whereas the generic expectation for the shape of the three-point function in single-field models with a small speed of sound is equilateral, i.e. a maximal signal in equilateral triangles, at special values of $\tilde{c}_3$ (around $-5$) the non-Gaussian signal peaks in enfolded (or collinear) triangles instead. In fact, the orthogonal template put forward in \cite{SSZ2009} to analyze the CMB data is closely related to an earlier proposal in \cite{Meerburg} designed with initial state modifications in mind, which also tends to peak in enfolded triangles. It would be interesting to find a specific Lagrangian function $P(X, \phi)$ that can lead to these special values of $\tilde{c}_3$, which might be impossible. 
Subsequently we will base our three-point function calculations on the more ``traditional" general single-field action described by a function $P(X, \phi)$, which can always be mapped to the corresponding Goldstone boson action using (\ref{GBP}).

A specific example of much interest is the string-inspired model of DBI inflation \cite{AST:UVDBI, Chen:IRDBI},
\[ P(X,\phi)=-f(\phi)^{-1}\sqrt{1-2Xf(\phi)}+f(\phi)^{-1}-V(\phi) . \]
In this case, 
\begin{eqnarray*}
\Sigma & = & \frac{H^{2} M_p^2 \, \epsilon}{c_{s}^{2}}, \\
\lambda & = & \frac{H^{2} M_p^2 \, \epsilon}{2c_{s}^{4}}(1-c_{s}^{2}).
\end{eqnarray*}
Substituting this into (\ref{GBP}) one finds the corresponding $\tilde{c}_{3} =  \frac{3}{2}(1-c_{s}^{2})$, which equals $3/2$ in the small sound speed limit. This will cause the leading contribution in the three-point function to vanish for DBI models, which will be of importance when 
comparing the theoretical non-Gaussian amplitudes and shapes to the observational constraints. 

In order to use the action given by (\ref{eq:action}) to calculate our quantity of interest, the three-point function, we need to determine the corresponding interaction Hamiltonian for the gauge-invariant perturbation variable. After first perturbing in the inflaton field $\phi=\phi_{0}+\delta\phi$, we then convert this quantity to the gauge-invariant fluctuation $\zeta=-(H/\dot{\phi}_{0})\delta\phi$.  Expanding the action to third order in these perturbations, the effective interaction Hamiltonian associated with the cubic action is 
\cite{Chen:2005fe}
\begin{eqnarray}
H_{I}(\eta) & = & -\int d^{3}x\left[-a \left( \Sigma ( 1-\frac{1}{c_{s}^{2}})+2\lambda \right) \frac{(\dot{\zeta}_{x}){}^{3}}{H^{3}}+\frac{a^{2}\epsilon}{c_{s}^{4}}(\epsilon-3+3c_{s}^{2})\zeta_{x}\dot{\zeta}_{x}^{2}\right.\nonumber \\
&  & \left.+\frac{a^{2}\epsilon}{c_{s}^{2}}(\epsilon-2s+1-c_{s}^{2})\zeta_{x}(\partial\zeta_{x})^{2}-2\frac{\epsilon}{c_{s}^{2}}a\dot{\zeta}{}_{x}(\partial\zeta_{x})(\partial\chi)\right],\label{eq:interH}
\end{eqnarray}
where overdots represent derivatives with respect to conformal time. This interaction Hamiltonian will serve as our starting point for the perturbative calculation of the three-point function in a modified initial state.

\subsection{The three-point function}

We are interested in calculating the three-point correlation function $\langle\zeta_{k_{1}}\zeta_{k_{2}}\zeta_{k_{3}}\rangle$, where $\zeta_{k}$ is the gauge-invariant primordial density perturbation of comoving momentum $k$.  To first order in the interaction Hamiltonian $H_{I}$ this correlation can be written as an integral over the free-field correlator $\langle\zeta_{k_{1}}\zeta_{k_{2}}\zeta_{k_{3}}H_{I}(\eta)\rangle$, 
\begin{eqnarray}
\label{eq:3ptfunction}
\langle\psi(\eta)|\zeta_{k_{1}}(\eta)\zeta_{k_{2}}(\eta)\zeta_{k_{3}}(\eta)|\psi(\eta)\rangle & = & \langle\psi(\eta_{0})|\zeta_{k_{1}}(\eta)\zeta_{k_{2}}(\eta)\zeta_{k_{3}}(\eta)|\psi(\eta_{0})\rangle \\
&  & \hspace{-0.5in} -i\int_{\eta_{0}}^{\eta}d\eta'\langle\psi(\eta_{0})|[\zeta_{k_{1}}(\eta)\zeta_{k_{2}}(\eta)\zeta_{k_{3}}(\eta),H_{I}(\eta')]|\psi(\eta_{0})\rangle+\mathcal{O}(H_{I}^{2}), \nonumber
\end{eqnarray}
where the time-evolved initial state is given by
\[  |\psi(\eta)\rangle = Te^{-i\int_{\eta_0}^\eta H_I(\eta') d\eta'}|\psi(\eta_0)\rangle. \]
The first term on the right in eq. \eqref{eq:3ptfunction} will be zero if the initial state $|\psi(\eta_{0})\rangle$ is Gaussian, as we will assume. 
The interaction Hamiltonian will encode all higher-order contributions in the effective action, so it will be sufficient to consider interactions to lowest order in
$H_{I}$.  Recall that if the initial state is assumed to be BD at the onset of inflation, the 3-point correlator at tree level is usually computed from $\eta_{0} \rightarrow-\infty$.  In our case we assume there is some part of the physics of which we are ignorant, ending up in the initial state at the start of inflation.  Consequently we integrate from some initial time $\eta_{0}$ representing the initial time corresponding to the cutoff scale in physical momenta $p = k/a$.
To stress this again, $\eta_0$ will therefore be $k$-dependent and lead to a scale-invariant spectrum, which we require for our analysis. 
The presence of such a cutoff must be explicit because initial state modifications would otherwise have infinite time to interact, and the 3-point correlator would then be infinitely large - and hence excluded by observation!  In our case it is the time at which we chose to set the initial state to be Gaussian-distributed parameterized by an \emph{a priori} arbitrary complex Bogolyubov parameter. Relating the initial time to the moment the physical momentum of a mode reaches the cutoff scale implies that all modes have equal time to evolve until they exit the horizon, which allows for the preservation of scale-invariance in the bispectrum. 

One can expand the second line of the three-point correlator (\ref{eq:3ptfunction}) into a product of Wightman functions $G_{k}^{>}$.  These are defined as 
\begin{equation}
\langle\zeta_{k_{1}}(\eta) \zeta_{k_{2}} (\eta') \rangle = (2\pi)^{3}\delta^{(3)}(\vec{k}_{1}+\vec{k}_{2})G_{k_{1}}^{>}(\eta,\eta').\label{G2}
\end{equation}
The Wightman functions can be found by solving the classical equations of motion of the inflaton field action eq. \eqref{eq:action} minimally coupled to gravity,
\[ u_{k}(\eta) = \frac{iH}{\sqrt{4\epsilon c_{s}k^{3}}}(1+ikc_{s}\eta)e^{-ikc_{s}\eta}. \]
The quantized form of the dimensionless curvature perturbation is given by
\[ \zeta_{k}(\eta) = u_{k}(\eta)a_{k}+u_{-k}^{*}(\eta)a_{-k}^{\dagger}, \]
where $a_{k}$ and $a_{k}^{\dagger}$ are the annihilation and creation
operators respectively and the mode-functions $u_k(\eta)$ are dimensionless by introducing the appropriate powers of the 
reduced Planck mass $M_p$. Using this form of $\zeta_{k}(\eta)$ one obtains the Wightman function
\[ G_{k}^{>}(\eta,\eta') = \frac{H^{2}}{4 M_p^2 \epsilon c_{s} k^3}(1+ikc_{s}\eta)(1-ikc_{s}\eta')e^{-ikc_{s}(\eta-\eta')}. \]
We compute the late-time cases of interest to us,
\[ G_{k}^{>}(0,\eta) = \frac{H^{2}}{4 M_p^2 \epsilon c_{s} k^3}(1-ikc_{s}\eta)e^{ikc_{s}\eta}, \]
and
\begin{equation}
\partial_{\eta}G_{k}^{>}(0,\eta)=-\frac{H}{4 M_p^2 \epsilon k}\frac{c_{s}}{a(\eta)}e^{ikc_{s}\eta},\label{eq:derG>}
\end{equation}
with $a(\eta)=-1/\eta H$ during inflation in the usual assumption $\dot{H}\simeq0$.  To explicitly compute the correlation (\ref{eq:3ptfunction}), a useful identity is
\begin{eqnarray}
\langle\zeta_{k_{1}}(\eta)\zeta_{k_{2}}(\eta)\zeta_{k_{3}}(\eta)\rangle & \sim & -i\int_{\eta_{0}}^{\eta}d\eta'\langle\psi(\eta_{0})|[\zeta_{k_{1}}(\eta)\zeta_{k_{2}}(\eta)\zeta_{k_{3}}(\eta),H_{I}(\eta')]|\psi(\eta_{0})\rangle\nonumber \\
& = & -2\mathcal{R}e\left[\int_{\eta_{0}}^{\eta}id\eta'\langle\psi(\eta_{0})|\zeta_{k_{1}}(\eta)\zeta_{k_{2}}(\eta)\zeta_{k_{3}}(\eta)H_{I}(\eta')|\psi(\eta_{0})\rangle\right],\label{eq:threepoint1}
\end{eqnarray}
which removes the commutator and assures that we are dealing with a true observable. Next we can express $\langle\zeta_{k_{1}}\zeta_{k_{2}}\zeta_{k_{3}}\rangle$
in terms of these Wightman functions. For illustrational purposes let us just consider the first term on the right in eq. \eqref{eq:interH} in the absence of initial state modifications. The three-point function at late times $\eta \rightarrow 0$ is then
\begin{eqnarray*}
\langle\zeta_{k_{1}} (0) \zeta_{k_{2}} (0) \zeta_{k_{3}} (0) \rangle_{\mathrm{BD}} & = & -12(2\pi)^{3}\delta^{(3)}(\sum\vec{k}_{i}) \left[ \Sigma \left(1-\frac{1}{c_{s}^{2}} \right)+2\lambda \right] \frac{1}{H^{3}}\times\\
&  & \mathcal{R}e\left[i\int_{\eta_{0}}^{0}d\eta' \ a\partial_{\eta'}G_{k_{1}}^{>}(0,\eta')\partial_{\eta'}G_{k_{2}}^{>}(0,\eta')\partial_{\eta'}G_{k_{3}}^{>}(0,\eta')\right].
\end{eqnarray*}
The factor 12 comes from $2$ times the real part and the 6 possible permutations, and we have added the subscript to indicate that this is the result for the Bunch-Davies vacuum.  From here on we will omit the $\eta=0$ dependence of the correlation function, since we are only computing it at late times.  Using the Wightman functions \eqref{eq:derG>} we obtain
\begin{equation}
\hspace{-0.3in} \langle\zeta_{k_{1}}\zeta_{k_{2}}\zeta_{k_{3}}\rangle_{\mathrm{BD}} = \frac{12}{8}(2\pi)^{3}\delta^{(3)} ( \sum \vec{k}_{i}) \left[ \Sigma \left( 1-\frac{1}{c_{s}^{2}} \right)+2\lambda \right] \frac{c_{s}^{3}}{8 M_p^6 \epsilon^3}\mathcal{R}e\left[i\int_{\eta_{0}}^{0}\frac{d\eta'}{a^{2}}\frac{1}{k_{1}k_{2}k_{3}}e^{ik_{t}c_{s}\eta'}\right],\label{eq:zetazetezeta}
\end{equation}
where $k_{t}=k_{1}+k_{2}+k_{3}$.  In the limit $\eta_{0}\rightarrow-\infty$, the (analytically continued) integral becomes
\begin{eqnarray*}
\frac{H^{2}}{k_{1}k_{2}k_{3}}\int_{-\infty}^{0}d\eta' \ \eta'^2 e^{ik_{t}c_{s}\eta'} & = & -\frac{H^{2}}{k_{t}^{3}k_{1}k_{2}k_{3}}\frac{2}{ic_{s}^{3}}.
\end{eqnarray*}
Substituting this into eq. \eqref{eq:zetazetezeta} we find
\begin{eqnarray}
\langle\zeta_{k_{1}}\zeta_{k_{2}}\zeta_{k_{3}}\rangle_{BD} & = & -3(2\pi)^{3}\delta^{(3)} ( \sum\vec{k}_{i}) \left[ \Sigma \left(1-\frac{1}{c_{s}^{2}} \right)+2 \lambda \right] \frac{H^{2}}{8 M_p^6 \epsilon^3}\frac{1}{k_{1}k_{2}k_{3}k_{t}^{3}}.\label{eq:zetzetzet}
\end{eqnarray}
To relate this to the familiar result in Chen \emph{et al.} \cite{Chenetal} one uses the fact that 
$\Sigma=\epsilon M_p^2 H^{2}/c_{s}^{2}$, leading to the following expression:
\begin{equation}
\langle\zeta_{k_{1}}\zeta_{k_{2}}\zeta_{k_{3}}\rangle_{\mathrm{BD}}=-\frac{3}{8}(2\pi)^{3}\frac{H^{4}}{M_p^4 \epsilon^{2}c_{s}^{2}} \left( 1-\frac{1}{c_s^2}+\frac{2\lambda}{\Sigma} \right) \delta^{(3)}(\sum\vec{k}_{i})\frac{1}{k_{1}k_{2}k_{3}k_{t}^{3}}.
\label{eq:contr1}
\end{equation}
As is well-known, the momentum dependence of this contribution to the three-point function is well-approximated by the equilateral template, i.e. its magnitude maximizes in equilateral triangles. The full three-point function obviously contains additional contributions from the other terms in the interaction Hamiltonian. As mentioned earlier, in DBI models the above contribution vanishes exactly and another leading order equilateral term dominates (in the limit of slow-roll and small sound speed). All other contributions turn out to be higher order in the slow-roll expansion \cite{Chenetal}.

\subsection{Bispectrum corrections due to a modified vacuum state}
Next consider the following Bogolyubov transformation of the initial state,
\begin{eqnarray}
u_{k} & \rightarrow  & \alpha_{k}v_{k}+\beta_{k}v_{k}^{*},
\end{eqnarray}
with $|\alpha_k|^2-|\beta_k|^2=1$. We will be interested in small departures from the usual BD vacuum, so  $| \beta_{k}|/|\alpha_{k}| \ll1$, as required by power spectrum constraints.  For the BD vacuum ($\alpha_k \equiv 1$), our results will of course reduce to (\ref{eq:zetzetzet}).  Now let us suppose $\beta_{k}\neq0$ and compute the corresponding bispectrum to lowest order in $\beta_{k}$.  This will result in one of the modes effectively having negative energy and will carry along a corresponding factor of $\beta_{k}$.  As emphasized previously, we will have to introduce an initial time cutoff since we expect particle interaction before the end of inflation to dominate the bispectrum, which could become arbitrarily large if the earliest modes have infinite time to grow (before they cross the horizon). To preserve scale-invariance, we introduce a fixed physical cutoff scale and determine the intial time as $k c_s \eta_{0}= \Lambda_c/H^*$. Of course, scale-invariance also requires that the complex Bogolyubov parameter depends at most only weakly upon the comoving momentum $k$. The starting point will therefore be a constant complex parameter $\beta \equiv |\beta| \exp{(i \delta)}$ that parameterizes the departure from the BD state. Under these assumptions, the correction to the three-point correlation function eq.(\ref{eq:contr1}) is
\begin{eqnarray}
\langle\zeta_{k_{1}}\zeta_{k_{2}}\zeta_{k_{3}}\rangle_{\mathrm{nBD}} & = & \frac{4}{8}(2\pi)^{3}\delta^{(3)}(\sum\vec{k}_{i})\left[ \Sigma \left(1-\frac{1}{c_{s}^{2}} \right)+2\lambda \right] \frac{H^{6}}{\dot{\phi}_{0}^{6}}c_{s}^{3}\times \nonumber\\
&&\mathcal{R}e\left[\sum_{j}^{3}i\beta_{k_{j}}\int_{\eta_{0}}^{0}\frac{d\eta'}{a^{2}}\frac{1}{k_{1}k_{2}k_{3}}e^{i\tilde{k}_{j}c_{s}\eta'}\right],\label{eq:zetazetawinimod}\end{eqnarray}
where $\tilde{k}_{j}=k_{t}-2k_{j}$.  The integral is evaluated to be equal to 
\begin{equation}
\frac{H^{2}}{k_{1}k_{2}k_{3}}\sum_{j}\int_{\eta_{0}}^{0}d\eta' \ \eta'^{2}e^{i\tilde{k}_{j}c_{s}\eta'} = \frac{H^{2}}{k_{1}k_{2}k_{3}}\sum_{j}\left[i\frac{2}{\tilde{k}_{j}^{3}c_{s}^{3}}+i\frac{e^{i\tilde{k}_{j}c_{s}\eta_{0}}}{\tilde{k}_{j}c_{s}}\left(\eta_{0}^{2}+i\frac{2\eta_{0}}{\tilde{k}_{j}c_{s}}-\frac{2}{\tilde{k}_{j}^{2}c_{s}^{2}}\right)\right].\label{eq:integral2}
\end{equation}
The first term is similar to our previous result, with the substitution of $\tilde{k}_{j}\leftrightarrow k_{t}$.  
The remaining terms are a result of the presence of a cutoff. The final result is
\begin{eqnarray}
\langle\zeta_{k_{1}}\zeta_{k_{2}}\zeta_{k_{3}}\rangle_{\mathrm{nBD(1)}} & = & -\frac{1}{4}(2\pi)^{3}\frac{H^{4}|\beta|}{\epsilon^{2}c_{s}^{4}}\left(c_{s}^{2}-1+\frac{2\lambda c_{s}^{2}}{\Sigma}\right)\delta^{(3)}(\sum\vec{k}_{i})\frac{1}{k_{1}k_{2}k_{3}}\times\\\nonumber
&  & \hspace{-0.5in} \sum_{j}\left[\frac{c_{s}^{2}\eta_{0}^{2}}{2}\frac{\cos(\tilde{k}_{j}c_{s}\eta_{0}+\delta)}{\tilde{k}_{j}}-c_{s}\eta_{0}\frac{\sin(\tilde{k}_{j}c_{s}\eta_{0}+\delta)}{\tilde{k}_{j}^{2}}+\frac{\cos \delta-\cos(\tilde{k}_{j}c_{s}\eta_{0}+\delta)}{\tilde{k}_{j}^{3}}\right]\label{bd1a}.
\end{eqnarray}
Let us pause for a moment to interpret this expression. First, we notice that all terms contain a phase shift $\delta$ as a result of the phase in the Bogolyubov parameter. As will become clear this phase significantly influences the projection onto the observational templates and therefore the 
constraints on the vacuum state modification that can be derived. 

Secondly, we notice that in the limit $\tilde{k}_{j}\rightarrow0$ (and unlike previous
results \cite{Meerburg}) all terms scale as $1/\tilde{k}_{j}$, which might cause some concern when considering 
the limit $\tilde{k}_{j} \rightarrow 0$. We certainly expect the result to be finite for any phase $\delta$ and any combination 
of the three comoving momenta and fortunately explicit computation of the limit $\tilde{k}_j\rightarrow 0$ in the appendix 
confirms this expectation. The first maximum occurs for $\tilde{k}_j \propto 1/c_s \eta_0$ and since $k c_s \eta_0 \gg 1$, where $k$ 
would be the largest momentum scale under consideration, this effectively corresponds to the enfolded limit $\tilde{k}_j/ k \sim 0$. 
Also note that this localized form of enhancement would scale as $(k c_s \eta_0)^3$, which is expected to be very large.  Projection onto observational
templates generically results in the loss of at most one factor of $k c_s \eta_0$, which would then provide us with a quadratic enhancement in 
the large number $k c_s \eta_0$. A detailed analysis of the projections and constraints will be discussed in section 4.   

After this rather explicit introduction and discussion of the corrected three-point function due to the first term in the interaction Hamiltonian 
we would now like to present the complete (leading order) result for the corrected three-point function, for which the detailed calculations 
are presented in the appendix. Using the expression for the power spectrum of general single-field inflation
\begin{eqnarray}
P_{\zeta}&=& \frac{1}{8\pi^{2}}\frac{H^{2}}{c_{s}\epsilon},
\end{eqnarray}
we can write the leading order behavior, to first order in $\beta$, as follows
\begin{eqnarray}
\langle\zeta_{k_{1}}\zeta_{k_{2}}\zeta_{k_{3}}\rangle & = & (2\pi)^{7}|\beta]P_{\zeta}^{2}\delta(\sum\vec{k}_{i})\frac{1}{k_{1}^{3}k_{2}^{3}k_{3}^{3}}\times\nonumber\\
&  & \left[\sum_{j} \left( \frac{1}{c_{s}^{2}}-1+\frac{2\lambda}{\Sigma} \right) (k_{1}c_{s}\eta_{0})^{3}\mathcal{B}_{k_{j}}^{(1)}+(k_{1}c_{s}\eta_{0})^{2} \left( \left( \frac{1}{c_{s}^{2}}-1\right) \left( \mathcal{B}_{k_{j}}^{(2)}+\mathcal{B}_{k_{j}}^{(3)} \right)+\right.\nonumber \right. \\
&  & \left. \left. +\frac{\epsilon}{c_{s}^{2}} \left( \mathcal{B}_{k_{j}}^{(2)}+\mathcal{B}_{k_{j}}^{(3)}-\mathcal{B}_{k_{j}}^{(4)} \right)-\frac{2s}{c_{s}^{2}}\mathcal{B}_{k_{j}}^{(3)} \right)+\right.\nonumber\\
&  & \left. k_{1}c_{s}\eta_{0} \left( \frac{-2s+1-c_{s}^{2}}{c_{s}^{2}}\mathcal{B}_{k_{j}}^{(5)}+\frac{\epsilon}{c_{s}^{2}} \left( \mathcal{B}_{k_{j}}^{(5)}-\mathcal{B}_{k_{j}}^{(6)} \right) \right) +...\right],
\end{eqnarray}
where the $...$ represent terms further suppressed slow-roll and in the cutoff $\eta_0$. Here the different functions $\mathcal{B}$ equal
\begin{eqnarray}
\mathcal{B}_{k_{j}}^{(1)} &=& \frac{k_{2}^{2}k_{3}^{2}}{k_{1}}\left[\frac{1}{2}S_{1} \left(1-\frac{2}{(\tilde{k}_{j}c_{s}\eta_{0})^{2}} \right)-\frac{S_{2}}{\tilde{k}_{j}c_{s}\eta_{0}}+\frac{\cos \delta }{(\tilde{k}_{j}c_{s}\eta_{0})^{3}}\right],\label{eq:B1} \\
\mathcal{B}_{k_{j}}^{(2)} & = & \frac{3}{4}k_{2}^{2}k_{3}^{2}\left[\left(\frac{\tilde{k}_{j}-k_{j}}{k_{j}^{2}}+\frac{\tilde{k}_{j}+k_{j+1}}{k_{j+1}^{2}}+\frac{\tilde{k}_{j}+k_{j+2}}{k_{j+2}^{2}}\right)\left(\frac{\cos \delta }{(\tilde{k}_jc_s\eta_0)^2}-\frac{S_{1}}{\tilde{k}_{j}c_{s}\eta_{0}}\right)\right.\nonumber\\&&\left.+\left(\frac{1}{k_{j}}-\frac{1}{k_{j+1}}-\frac{1}{k_{j+2}}\right)S_{2}\right],\label{eq:B2}\\
\mathcal{B}_{k_{j}}^{(3)} & = & \frac{1}{8}\left[\frac{k_{2}k_{3}}{k_{1}}(2k_{j}^{2}-k_{j+1}^{2}-k_{j+2}^{2})S_{2}-\right.\nonumber\\
&& \left.\frac{1}{k_1^2}\left(k_1k_2k_3+\tilde{k}_j(k_jk_{j+1}+k_jk_{j+2}-k_{j+1}k_{j+2}) \right) \left( \frac{\cos \delta }{(\tilde{k}_{j}c_{s}\eta_{0})^{2}}-\frac{S_{1}}{\tilde{k}_{j}c_{s}\eta_{0}}\right)\right]\label{eq:B3}, 
\end{eqnarray}

\begin{eqnarray}
\mathcal{B}_{k_{j}}^{(4)} & = & \frac{1}{4}\frac{1}{k_{1}^{2}}\left(k_{j}^{4}(k_{j+1}+k_{j+2})+k_{j+1}^{4}(k_{j+2}-k_{j})+k_{j+2}^{4}(k_{j+1}-k_{j})+\right.\nonumber\\&&\left.+k_{j}^{2}k_{j+2}^{2}(k_{j}-2k_{j+1}-k_{j+2})+k_{j+1}^{2}k_{j+2}^{2}(2k_{j}-k_{j+1}-k_{j+2})+\right.
\nonumber\\&&\left.k_{j}^{2}k_{j+1}^{2}(k_{j}-k_{j+1}-2k_{j+2})\right)\left(\frac{\cos \delta }{(\tilde{k}_{j}c_{s}\eta_{0})^{2}}-\frac{S_{1}}{\tilde{k}_{j}c_{s}\eta_{0}}-S_{2}\right),\label{eq:B4}\\
\mathcal{B}_{k_{j}}^{(5)} & = & \frac{3}{8}\frac{(k_{1}^{2}+k_{2}^{2}+k_{3}^{2})}{k_{1}}(k_{j}k_{j+1}+k_{j}k_{j+2}-k_{j+1}k_{j+2})S_{1},\label{eq:B5}\\
\mathcal{B}_{k_{j}}^{(6)} & = & \frac{1}{4}\frac{1}{k_{1}}\left(\frac{1}{2}(k_{1}^{4}+k_{2}^{4}+k_{3}^{4})-(k_{1}^{2}k_{2}^{2}+k_{2}^{2}k_{3}^{2}+k_{3}^{2}k_{1}^{2})\right)S_{1},\label{eq:B6}\end{eqnarray}
and \begin{eqnarray*}
S_{1}(\tilde{k}_{j}c_{s}\eta_{0},\delta) & = & \frac{\cos(\tilde{k}_{j}c_{s}\eta_{0}+\delta)}{\tilde{k}_{j}c_{s}\eta_{0}},\\
S_{2}(\tilde{k}_{j}c_{s}\eta_{0},\delta) & = & \frac{\sin(\tilde{k}_{j}c_{s}\eta_{0}+\delta)}{\tilde{k}_{j}c_{s}\eta_{0}}.
\end{eqnarray*}
We have organized the bispectrum into pieces with different (decreasing) powers of $k_{1}c_{s}\eta_{0}$, i.e. the ratio between the Hubble radius and the largest physical momentum scale at the cutoff time $\eta_0$, i.e. $k_{1}c_{s}\eta_{0}=(k_1/a_0)/(H/c_s)$ which can be as large as $10^3$. For our purposes only the first two terms will be relevant as these are both enhanced in $k_{1}c_{s}\eta_{0}$. From here on we will refer to these terms as 
the leading (proportional to $(k_{1}c_{s}\eta_{0})^3$) and the subleading (proportional to $(k_{1}c_{s}\eta_{0})^2$) term, respectively. 

On first impression it seems that the momentum-dependence, or the shape, of the different terms in the above bispectrum 
strongly deviate from the constrained local, equilateral and orthogonal templates. Most importantly, all terms rapidly oscillate. 
Furthermore, as mentioned already, the leading-order term maximizes in the collinear limit.
In \cite{Meerburg} we proposed a template for collinear triangles, which unfortunately is only a marginal improvement as compared to the local and 
equilateral template. Ideally, one would like to introduce a template that can (also) incorporate oscillations with an \emph{a priori} undetermined frequency, introducing an additional parameter. It remains to be seen whether oscillations in the momentum dependence of the bispectrum are observable at all, 
as the finite number of pixels in the data might make determination of this additional frequency parameter very difficult.  We will not be concerned with this problem here since we will only be interested in projecting onto bispectrum templates that have already been constrained. 
We hope to report on the opportunities for the detection of oscillations in the primordial bispectrum in future work.

\section{Observational constraints}

\begin{figure}
\includegraphics[scale=0.58]{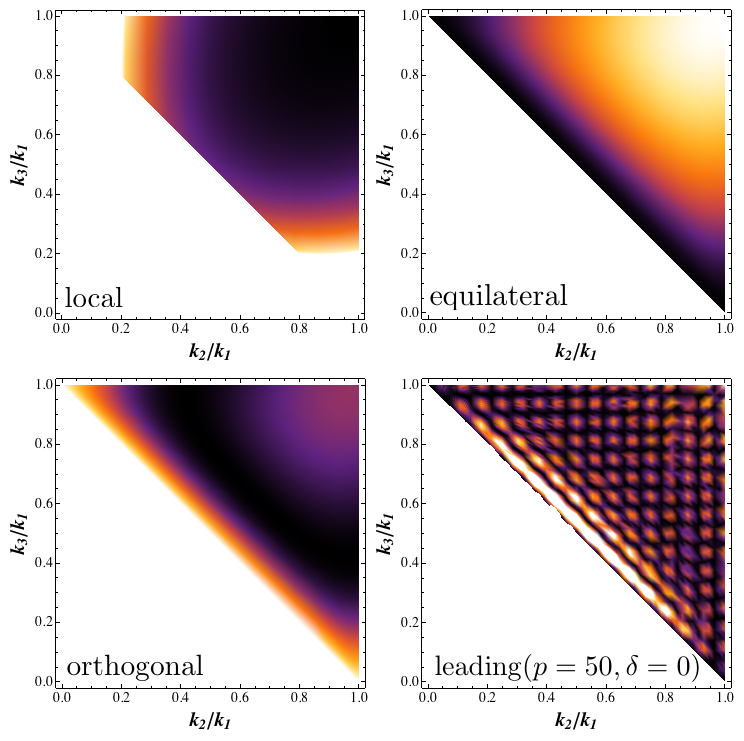}
\caption{Here we plotted the `density' of the different non-Gaussian shapes. As can be seen from the color (shading), large density corresponds to lighter color (shading). It is clear from inspection that the equilateral template has zero density on the collinear line (spanned by $x_{2}+x_{3}=1$), while the leading order term from initial state modifications maximizes in the vicinity of this line, resulting in virtually zero overlap (which is already small due to the oscillatory nature of the leading term).}
\label{fig:densities}
\end{figure}
\subsection{Preliminaries}
In order to use our results to place constraints on $|\beta|$ we need to relate them to existing templates.  This is done by computing the 
dot product between our bispectrum result and the local, equilateral and orthogonal templates, which are standardized and whose magnitude has 
been constrained in the literature \cite{WMAP5, SSZ2009}.  To facilitate this, let us first rewrite our result in terms of the dimensionless 
variables $x_{2}=k_{2}/k_{1}$ and $x_{3}=k_{3}/k_{1}$, scaling out a factor of $k_1^{-6}$, indicating that the bispectrum is 
scale-invariant, where the comoving momentum magnitude $k_1$ is identified as the largest momentum. Because of the triangle constraint 
the shape function then only depends on a finite (triangle) domain in $x_2$ and $x_3$ space ($0 \leq x_2 \leq 1$ and $1-x_2 \leq x_3 \leq 1$).  Using these dimensionless variables, the leading shape function is then expressed as 
\begin{eqnarray}
F_{\rm leading}(x_{2},x_{3},p,\delta) & = & \frac{1}{x_{1}x_{2}}\left[\frac{1}{2}\frac{\cos(p(x_{1}+x_{2}-1)+\delta)}{p(x_{1}+x_{2}-1)}-\frac{\sin(p(x_{1}+x_{2}-1)+\delta)}{p^{2}(x_{1}+x_{2}-1)^{2}}\right.\nonumber \\
&  & \left.+\frac{\cos\delta-\cos(p(x_{1}+x_{2}-1)+\delta)}{p^{3}(x_{1}+x_{2}-1)^{3}}+\mathrm{cyc. \ perm.}\right],\label{eq:fleading}\end{eqnarray}
where $p \equiv k_{1}c_{s}\eta_{0}$ is the enhancement factor we encountered several times before. Having identified the shape of the leading 
contribution we automatically also fixed a particular expression for the non-Gaussian amplitude $f_{\mathrm{NL}}$. It is important to
note that the above shape function implies that a factor of $p^{3}$ has been incorporated in the non-Gaussian amplitude $f_{\mathrm{NL}}$.

There are three different observational templates that we will be projecting our signal onto.  These are the \emph{squeezed} or \emph{local}, \emph{equilateral} and \emph{orthogonal} templates. In terms of $x_2$ and $x_3$, their shape functions are expressed as follows:
\begin{eqnarray}
F_{\rm local}(1,x_{2},x_{3}) &=& 2\left(\frac{1}{x_{2}^{3}}+\frac{1}{x_{3}^{3}}+\frac{1}{x_{2}^{3}x_{3}^{3}}\right), \\
F_{\rm eq}(1,x_{2},x_{3})&=&6\left(-\frac{F_{\rm local}}{2}-\frac{2}{x_{2}^{2}x_{3}^{2}}+\frac{1}{x_{2}^{2}x_{3}^{3}}+\frac{1}{x_{2}x_{3}^{3}}+\frac{1}{x_{2}x_{3}^{2}}+\mathrm{perm.}\right), \\
F_{\rm ort}(1, x_2, x_3)&=&3 \left( F_{\rm eq}-\frac{4}{x_2^2 x_3^2} \right) .
\end{eqnarray}
Each of these templates has a different physical origin. The squeezed template was introduced by Spergel and Komatsu \cite{SperKom} to parameterize non-Gaussianities resulting from a quadratic correction in real space (i.e. a local term), corresponding to the leading nonlinear coupling between the inflaton field and the curvature perturbation at superhorizon scales.  The equilateral template was introduced by Creminelli \emph{et al.} \cite{BCZ2004} to parameterize nonlinear higher-derivative corrections in the inflaton field, whose nonlinear magnitude is fixed at horizon crossing.  As mentioned, 
the orthogonal template was recently introduced by Senatore \emph{et al.} \cite{SSZ2009} to allow for an optimal comparison with exotic 
non-canonical single-field scenarios that are, rather surprisingly, not of the equilateral type. A closely related template was also put forward in 
earlier work \cite{Meerburg}. 

In order to make comparisons between these different shapes, in \cite{BCZ2004, Fergusson} dot products, cosines (the normalized dot product) and projection factors (to be introduced later) were defined to be
\begin{eqnarray}
F_{X}\star F_{Y} & \equiv & \int_{\Delta_{k}}dk_{1}dk_{2}dk_{3}\frac{k_{1}^{4}k_{2}^{4}k_{3}^{4}}{k_{t}}F_{X}(k_{1},k_{2},k_{3})F_{Y}(k_{1},k_{2},k_{3}),\nonumber \\
\mathrm{cos}(F_{X},F_{Y}) & \equiv & \frac{F_{X}\star F_{Y}}{(F_{X}\star F_{X})^{1/2}(F_{Y}\star F_{Y})^{1/2}} .
\label{eq:correlator}
\end{eqnarray}
In the case where both $F_{Y}$ and $F_{X}$ are scale-invariant we can remove the integral over one variable, (i.e. $k_{1}$). 
In a previous definition of the dot (or star) product \cite{BCZ2004} the factor $1/k_t$ was omitted. It was argued in \cite{ Fergusson} that
the addition of this extra `weight' reduces the difference between the three-dimensional dot product for the bispectra and the two-dimensional dot product for the respective multipole equivalents. Since one actually measures the three-point function in two-dimensional multipole space, it is the 
two-dimensional dot product that one is interested in computing to derive constraints. Since the two-dimensional multipole bispectrum and 
its corresponding dot product are computationally challenging (and analytically ill-understood), it would indeed be a great improvement 
if a three-dimensional dot product could be introduced that guarantees reproduction of the two-dimensional dot product in multipole space. 
This is what the above dot product (including the $1/k_t$) is supposed to do, removing the need for a fully fledged two-dimensional analysis.  

We will take this opportunity to clarify why the assumption of scale-invariance produces a difference with the old dot product which is (approximately) 
only a factor of 1/2. Note that the weight function $1/k_t$ is given by $1/(1+x_{2}+x_{3})$ (after scaling out $k_1$), and the integration limits 
are given by $0\leq x_{2}\leq1$ and $1-x_{2}\leq x_{3}\leq1$.  Consequently $1+x_{2}\leq1+x_{2}+x_{3}\leq2+x_{2}$ or $1\leq1+x_{2}+x_{3}\leq3$, 
resulting on average in $k_t = 1+x_{2}+x_{3} \approx 2$, explaining the factor 1/2 difference. 
When computing the projection factor, a ratio of dot products that we are ultimately interested in, 
this factor 1/2 drops out and one ends up with exactly the same result using the former definition of the projection factor. 
We should emphasize that this is only true under the assumption of scale-invariance. Once this assumption is dropped one cannot 
scale out the $k_1$ explicitly and one has to perform the full three-dimensional integration in $k$-space. 
Since the dot product without the weight $1/k_t$ can be computed analytically we will actually be working with the standard definition
of the three-dimensional dot product (see appendix eq. \eqref{eq:dotproduct_old}) and assume the results will (approximately) equal the two-dimensional multipole space dot product.

Since we expect our bispectra to have poor overlap with the templates, and the dot product between the theoretically predicted bispectra is very hard to compute analytically, we 
do not consider the cosine between the spectra. However, to get an indication how well (or how poor) a template does as compared to each other, we can define the ratio between two cosines
\begin{eqnarray}
\mathcal{R}(X,Y)&\equiv& \left|\frac{\cos (F_X,F)}{\cos(F_Y,F)}\right|=\left|\frac{F_X\cdot F}{F_Y\cdot F}\right|\frac{\sqrt{F_Y\cdot F_Y}}{\sqrt{F_X\cdot F_X}},
\end{eqnarray}
where $F$ represents the theoretically predicted bispectrum, and $F_X$ and $F_Y$ represent different bispectrum templates. If $\mathcal{R}(X,Y)>1$, $F_X$ is a better template and if $\mathcal{R}(X,Y)<1$ the opposite is true.

Figure 2 shows the density of each shape which gives a qualitative indication of where the shapes peak as a function of the comoving momenta. We can immediately see that the orthogonal template naively appears to `best' match the qualitative features in our signal (at least for $\delta=0$), an intuition that will be (partially) supported by calculations of the dot product.  As an instructive example, let us consider the cross-correlations with the leading shape function eq.(\ref{eq:fleading}), which can be analytically determined,
\begin{eqnarray*}
F_{\rm local}  \cdot F_{\rm leading} & = & -\frac{3}{4}\frac{\cos\delta}{p}+\frac{\sin\delta}{p^2}+\mathcal{O}(p^{-3}),\\
F_{\rm eq} \cdot F_{\rm leading} & = & 2\frac{\sin\delta}{p^2}+\mathcal{O}(p^{-3}), \\
F_{\rm ort} \cdot F_{\rm leading} & = & \frac{1}{2}\frac{\cos\delta}{p}+3\frac{\sin\delta}{p^2}+\mathcal{O}(p^{-3}).
\end{eqnarray*}
Consequently both the dot product with the local as well as the orthogonal template peak for $\delta=0$, while the dot product 
with the equilateral template is $\pi/2$ out of phase. Note that we extracted three powers of $p$ into the amplitude, so the 
dot products should be multiplied with $p^3$ to determine the strength of the enhancement that remains. 
The reason for this reduction in the level of enhancement is simply that the original enhancement was localized in the 
(leading) three-point function and integrating over the full domain will smear out this enhancement, typically reducing it with 
one power of $p$. Clearly, the dot product with the equilateral template is additionally suppressed in $p$ as compared to the local 
and orthogonal template (for $\delta=0$ in fact all enhancement is lost). Visual inspection of the leading-order term in the density 
profile of figure \ref{fig:densities} indicates that the small dot product with the equilateral template is explained by the fact that this term 
peaks in the collinear limit, where the equilateral template actually vanishes. As mentioned before, this is precisely where the orthogonal 
template peaks, and therefore it is not a surprise that it would have a larger dot product. A significant feature of
the orthogonal template is that it is not positive-definite within the triangle domain. As such it shares a feature with the 
oscillating contributions, possibly resulting in an even larger dot product. For comparison we find $\mathcal{R}(F_{\mathrm{local}},F_{\mathrm{ort}})\sim 0.42$ for $\delta=0$ which 
quantitatively confirms the orthogonal template has the best overlap with the theoretical spectrum\footnote{As a reference, the overlap between the local and equilateral shape is $\cos(F_{\mathrm{local}},F_{\mathrm{eq}})\sim 0.41$.}. 

We conclude that for generic values of the 
phase $\delta$ ($\delta \neq \pi/2$), the local and orthogonal templates would provide the best constraints if the shape function eq.(\ref{eq:fleading}) 
would be a good approximation to the full three-point function.  As one should expect, the constraints 
resulting from these expressions will also depend on the tightness of the existing observational constraints on the corresponding $f_{\mathrm{NL}}$, 
which will favor the local template.

For completeness we should also consider, in addition to the term (\ref{eq:fleading}) which is a unique consequence of introducing a vacuum state modification, the non-Gaussian signature as a consequence of non-standard $c_{s}$ (i.e. $\mathcal O(\beta^0)$).  As computed in \cite{Chen:2005fe}, the leading terms are 
given by
\begin{eqnarray}
\mathcal{B}_{\lambda} & = & \frac{1}{k_{1}^{3}k_{2}^{3}k_{3}^{3}}\left(\frac{1}{c_{s}^{2}}-1-\frac{2\lambda}{\Sigma}\right)\frac{3k_{1}^{2}k_{2}^{2}k_{3}^{2}}{2k_{t}^{3}},\\
\mathcal{B}_{c} & = & \frac{1}{k_{1}^{3}k_{2}^{3}k_{3}^{3}}\left(\frac{1}{c_{s}^{2}}-1\right)\left(-\frac{1}{k_{t}}\sum_{i>j}k_{i}^{2}k_{j}^{2}+\frac{1}{2k_{t}^{2}}\sum_{i\neq j}k_{i}^{3}k_{j}^{3}+\frac{1}{8}\sum_{i}k_{i}^{3}\right),
\end{eqnarray}
where we neglected the corrections due to time-variation of various parameters as well as corrections to the solutions of the field equations of 
motion.  Both of these contributions have a $k$-dependence that strongly resembles the equilateral template,
\begin{eqnarray}
\mathrm{cos}(F_{\rm eq},\mathcal{B}_{\lambda}) \approx \mathrm{cos}(F_{\rm eq},\mathcal{B}_{c}) \approx 0.95.
\end{eqnarray}
This contribution is therefore expected to add a term to the equilateral non-Gaussian amplitude, in addition to the modified vacuum state effect. 
It will of course be interesting to see how these contributions compare.

\subsection{General inflationary models with a small speed of sound}

We can use the dot product, i.e. the correlator,  to define projections, which measures the relative ``leakage" of one shape function into one of
the available observational templates
\begin{equation}
\Delta(F_X,F_Y) \equiv \frac{F_{Y}\cdot F_{X}}{F_{X}\cdot F_{X}}.
\label{eq:projection}
\end{equation}
Using the normalizations (self-dot products) of the different templates that are given in the appendix, it is straightforward to compute 
the projection factors for the different templates with the leading shape behavior of the corrected three-point function due to a modified vacuum
state. For large $p$ this is well-described by the shape function that was discussed in the previous paragraph. The results are
\begin{eqnarray}
\Delta (F_{\rm leading},F_{\rm local}) & \approx & \frac{1}{176.5}\left[-\frac{3}{4}\frac{\cos \delta }{p}+\frac{\sin \delta }{p^2}\right],\label{eq:delta1}\\
\Delta (F_{\rm leading},F_{\rm eq}) & \approx & \frac{2}{7.9}\frac{\sin \delta }{p^2},\label{eq:delta2}\\
\Delta (F_{\rm leading},F_{\rm ort}) & \approx & \frac{3}{13.8}\left[\frac{1}{6}\frac{\cos \delta }{p}+\frac{\sin \delta }{p^2}\right],\label{eq:delta5}\\
\Delta (\mathcal{B}_{\lambda},F_{\rm eq}) & \approx & 0.01,\label{eq:delta3}\\
\Delta (\mathcal{B}_{c},F_{\rm eq}) & \approx & -0.05.\label{eq:delta4}\end{eqnarray}
From the corresponding amplitude of the three-point function, including the $p^3$ enhancement factor, one then derives the following 
expressions for the local, equilateral (including contributions due to a small speed of sound) and orthogonal non-Gaussian amplitudes:
\begin{eqnarray}
f_{\mathrm{NL}}^{\rm local} & \simeq & \frac{1}{176.5}\left(-\frac{5}{4}\cos \delta +\frac{5}{3}p^{-1}\sin \delta \right)\left(\frac{1}{c_{s}^{2}}-1+\frac{2\lambda}{\Sigma}\right)|\beta| p^{2}, \\
f_{\mathrm{NL}}^{\rm eq} & \simeq & -\frac{5}{3}\left[\left(\frac{1}{c_{s}^{2}}-1\right)\left(\frac{-2}{7.9}|\beta|p\sin \delta  +0.04\right)+\left(\frac{-2}{7.9} |\beta|p\sin \delta -0.01\right)\frac{2\lambda}{\Sigma}\right],\\
f_{\mathrm{NL}}^{\rm ort} & \simeq & \frac{1}{13.8}\left(\frac{5}{6}\cos \delta +5p^{-1}\sin \delta \right)\left(\frac{1}{c_{s}^{2}}-1+\frac{2\lambda}{\Sigma}\right)|\beta| p^{2}.
\end{eqnarray}
Again, in deriving these results we have assumed, based on the arguments presented earlier, that the three-dimensional dot products are good approximations of the two-dimensional dot products of the multipole bispectra. Also note that the two terms in the equilateral non-Gaussian 
amplitude that are not proportional to $|\beta|$ are a direct consequence of a small speed of sound, which produces an enhanced equilateral 
non-Gaussian component which cannot be neglected for $\delta = 0$.  The local and orthogonal contributions are, however, uniquely 
due to the modified vacuum state. We are now in a position to use these predictions to put constraints on the parameter space, consisting of 
$c_{s}$, $|\beta|$, $\delta$ and $p$.  The existing constraints on $f_{\mathrm{NL}}^{\rm local}$ \cite{Smith:2009jr}, $f_{\mathrm{NL}}^{\rm equil}$ \cite{WMAP5, SSZ2009} 
and $f_{\mathrm{NL}}^{\rm ort}$ \cite{SSZ2009} are given by 
\begin{eqnarray*}
-4 & \leq f_{\mathrm{NL}}^{\rm local}\leq & 80,\\
-125 & \leq f_{\mathrm{NL}}^{\rm equil}\leq & 435,\\
-369 & \leq f_{\mathrm{NL}}^{\rm ort} \leq & 71 .
\end{eqnarray*}
Based on these results let us now consider two extremes: $\delta=0$ (real $\beta$) and $\delta=\pi/2$ (imaginary $\beta$). 

For $\delta=0$, eq. \eqref{eq:delta2} vanishes, which implies the linearly enhanced equilateral contributions proportional to $|\beta|$ vanish. 
In that case we can actually use the equilateral constraints to provide a bound on the speed of sound, similar to that found by Senatore \emph{et al.} 
\cite{SSZ2009}, but as we will see this is not even required 
to derive an interesting constraint on $|\beta|$. One should also realize that the $\delta=0$ case describes the generic order of magnitude 
expectation, in the sense that as long as $\delta \neq \pi/2$ the local and orthogonal contributions remain $p^2$ enhanced, which is the most
important property as far as the constraints are concerned.
The constraints from the orthogonal template and the local template are very similar. Since 
$|\beta|$ is positive-definite this implies only the positive constraint remains (the negative constraint can be used for $\delta=\pi$). 
Note that the projection factor between the local and the leading shape function has
a negative sign, so the final constraint comes from the (negative) lower bound
on local non-Gaussianities $f_{\mathrm{NL}}^{\mathrm{local}}$, which is given by
\begin{eqnarray}
|\beta|&\lesssim& \frac{6 \times 10^2}{p^2} \left(\frac{1}{c_{s}^{2}}-1+\frac{2\lambda}{\Sigma}\right)^{-1},\;\;(\delta=0).
\end{eqnarray} 
To make this more concrete, let us assume that the ratio $2\lambda/\Sigma \ll 1/c_s^2$ in the small sound speed limit. 
The constraint then reads $|\beta| < 6 \times 10^2 c_s^2/p^2$ and after realizing that $p \equiv k_1 c_s \eta_0 = \Lambda_c c_s/H $ this 
implies the interesting, sound speed-independent, constraint $|\beta| < 6 \times 10^2 H^2/\Lambda_c^2$. This could easily give rise to a constraint 
on $|\beta|$ at the level of $10^{-4}$, independent of any specific model for $|\beta|$. In particular it 
seems to exclude Bogolyubov parameters linear in $H/\Lambda_c$ \cite{NPH} \footnote{Strictly speaking, starting with the assumption
that the Bobolyubov parameter is proportional to $\frac{H^\star}{\Lambda_c}$, which introduces another factor of $c_s$, one would obtain a lower 
bound on the ratio of $\frac{H}{\Lambda_c}$ depending on the speed of sound which turns out to be in conflict with constraints from the power spectrum 
in the limit of a small sound speed.}.  Therefore in the small speed of 
sound limit (and $2\lambda/\Sigma \ll 1/c_s^2$), we obtain an impressive constraint on $|\beta|$. As long as the phase $\delta \neq \pi/2$ 
this conclusion is unaffected. For values $\delta>\pi$ the sign in the leakage factors is altered and the best constraint is obtained from orthogonal non-Gaussianities. Due to the larger number bounds on this $f_{\mathrm{NL}}$, this results in a constraint that is one order of magnitude below the best constraint. 
\begin{figure}
\begin{centering}
\includegraphics[scale=0.55]{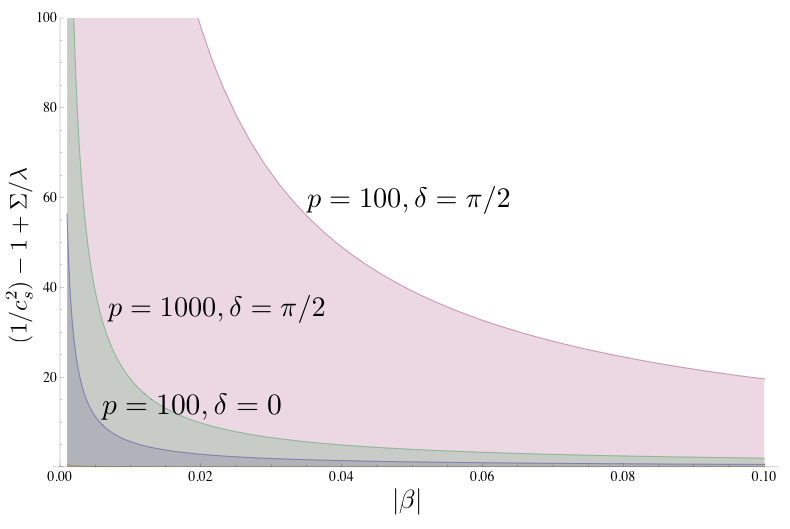}
\par\end{centering}

\caption{Constraining contours for different values of $p=c_{s}k_{1}\eta_{0}$ and two values of $\delta$. The constraints for intermediate values of $\delta$ should lie somewhere in between. For values $\pi/2<\delta<3\pi/2$ one has to consider the opposite (in sign) 
constraints on the various  $f_{\mathrm{NL}}$. The constraint for $p=1000$ and $\delta=0$ is too tight to be visible in this graph. 
For $\delta=\pi/2$ and $p=1000$ the constraint is still weaker than the constraint for $\delta=0$ and $p=100$.}
\label{fig:contours}
\end{figure}

Now let us briefly consider the special case $\delta=\pi/2$. For the local and orthogonal contributions this means the leading $p$ behavior 
is one order of magnitude less, which implies that the equilateral contribution is now of the same order. Looking at the equilateral contribution
we conclude that we can neglect the equilateral contributions from the zeroth order bispectrum (i.e. without initial state modification) as long
as $|\beta| > 0.1/p$. If the Bogolyubov parameter is smaller than that we can actually use the equilateral contribution to put a constraint on $c_s$. 
This time, the strongest constraint is derived from the orthogonal template and reads
\begin{eqnarray}
|\beta|&\lesssim& \frac{2 \times10^2}{p}\left(\frac{1}{c_{s}^{2}}-1+\frac{2\lambda}{\Sigma}\right)^{-1},\;\;(\delta=\pi/2).
\label{eq:leading_c2}
\end{eqnarray} 
Obviously, this constraint is much weaker and can be written as $|\beta| < 2 \times 10^2 H c_s/\Lambda_c$, in the limit of a small speed of sound
and $2\lambda/\Sigma \ll 1/c_s^2$. Since the enhancement in $p$ is reduced in this case one should worry about subleading 
contributions since they might be comparable in magnitude. Upon checking this one finds that for $\delta= \pi/2$ one should indeed take into account both the 
subleading and the leading order contribution. In the next section we will derive the projection factors for the subleading terms, since these are also
the relevant terms in DBI inflation for which $(1/c_{s}^2-1+2\Sigma/\lambda)$ vanishes identically.

\subsection{DBI models of inflation}
For DBI models (IR or UV) the leading-order term of the vacuum state corrected three-point function vanishes.
Consequently we will need to compute the dot product with the next-to-leading order terms. The next-to-leading contribution goes as 
(from eq.\eqref{eq:B2} and eq.\eqref{eq:B3}) 
\begin{eqnarray*} 
\langle\zeta_{k_1}\zeta_{k_2}\zeta_{k_3}\rangle_{sl} & \sim & \frac{1}{k_{1}^{3}k_{2}^{3}k_{3}^{3}}|\beta|\left(\frac{1}{c_{s}^{2}}-1\right)(k_{1}c_{s}\eta_{0})^{2}\sum_{j}(\mathcal{B}_{k_{j}}^{(2)}+\mathcal{B}_{k_{j}}^{(3)}),\end{eqnarray*}
where we have 2 distinguishable contributions,
\begin{eqnarray*}
\mathcal{B}_{k_{j}}^{(2)} & = & \frac{3}{4}k_{2}^{2}k_{3}^{2}\left[\left(\frac{\tilde{k}_{j}-k_{j}}{k_{j}^{2}}+\frac{\tilde{k}_{j}+k_{j+1}}{k_{j+1}^{2}}+\frac{\tilde{k}_{j}+k_{j+2}}{k_{j+2}^{2}}\right)\left(\frac{\cos \delta }{(\tilde{k}_jc_s\eta_0)^2}-\frac{S_{1}}{\tilde{k}_{j}c_{s}\eta_{0}}\right)\right.\\\nonumber&&\left.+\left(\frac{1}{k_{j}}-\frac{1}{k_{j+1}}-\frac{1}{k_{j+2}}\right)S_{2}\right],\label{eq:B2}\\
\mathcal{B}_{k_{j}}^{(3)} & = & \frac{1}{8}\left[\frac{k_{2}k_{3}}{k_{1}}(2k_{j}^{2}-k_{j+1}^{2}-k_{j+2}^{2})S_{2}-\right.\\
&& \left.\frac{1}{k_1^2}\left(k_1k_2k_3+\tilde{k}_j(k_jk_{j+1}+k_jk_{j+2}-k_{j+1}k_{j+2})\right)\left( \frac{\cos \delta }{(\tilde{k}_{j}c_{s}\eta_{0})^{2}}-\frac{S_{1}}{\tilde{k}_{j}c_{s}\eta_{0}}\right)\right]\label{eq:B3}. \\
\end{eqnarray*}
As was the case for the leading contribution, both terms are finite in the collinear limit (as shown explicitly in the appendix). 
Both contributions are of the same order of magnitude over the full triangle domain, so both have to be included in our analysis. 
One might argue that the cosine in both of these functions is suppressed and can be neglected. However, for an analytical treatment of the 
projection factors it turns out to be convenient to work with the full expression because of its finite collinear limit.   
Moreover, as we will see shortly, the phase $\delta$ significantly influences the final results for the projection factors. 

As before, let us rewrite these contributions in terms of the dimensionless variables $x_2$ and $x_3$. We will identify the shape of
the first sub-leading contribution as follows  
\begin{eqnarray}
{F}_{sl(1)} & = & -\frac{3}{4}\frac{1}{x_2 x_3}\left[\left(1-\frac{1}{x_2}-\frac{1}{x_3}\right)\frac{\sin(p(x_2+x_3-1)+\delta)}{p(x_2+x_3-1)}\right.\\\nonumber
&& +\left. \left(x_2+x_3-2 +\frac{1-2x_2+x_3}{x_2^2}+\frac{1+x_2-2x_3}{x_3^2}\right)\frac{\cos \delta -\cos(p(x_2+x_3-1)+\delta)}{p^2(x_2+x_3-1)^2}\right],
\label{eq:Fsl_1}
\end{eqnarray}
plus cyclic permutations. The shape of the second sub-leading term equals
\begin{eqnarray}
F_{sl(2)} & = & -\frac{1}{8}\frac{1}{x_2^3x_3^2}\left[x_2 x_3\left(1+x_2^2+x_3^2\right)\frac{\sin(p(x_2+x_3-1)+\delta)}{p(x_2+x_3-1)}\right.\\\nonumber
&& \hspace{-0.3in} -\left. \left((1+x_2^2+x_3^2)(x_2x_3+(x_2+x_3-1)(x_2+x_3-x_2x_3)\right)\frac{\cos \delta -\cos(p(x_2+x_3-1)+\delta)}{p^2(x_2+x_3-1)^2}\right],
\label{eq:Fsl_2}
\end{eqnarray}
plus cyclic permutations.  These identifications of the shape function imply we have extracted a factor of $p^2$ into the non-Gaussian 
amplitude.

\begin{figure}
\begin{centering}
\includegraphics[scale=0.45]{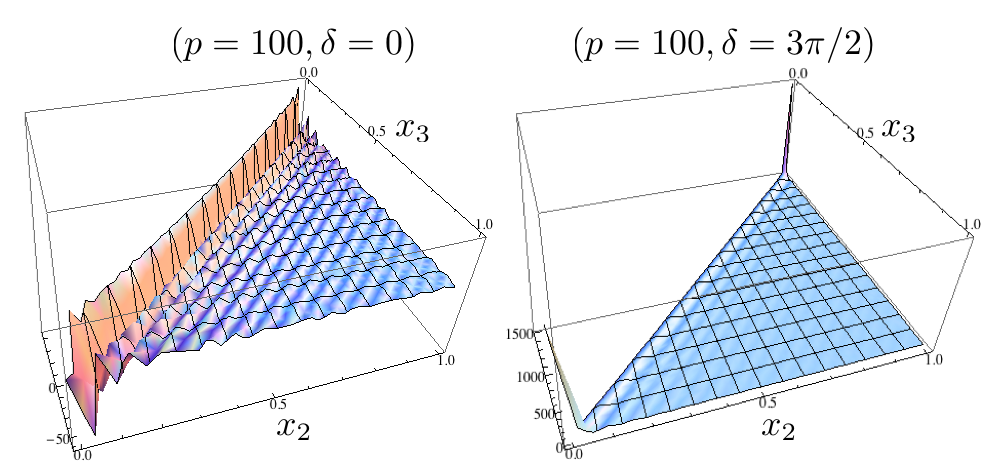}
\par\end{centering}
\caption{The shape of the first subleading contribution (eq.\protect\eqref{eq:Fsl_1}) for different values of $\delta$. For $\delta=0$ the shape of this function is very similar to the leading order term. Once $\delta\neq0$ the shape picks up a significant local feature which leads us to expect a significant overlap with the local non-Gaussian template. Since the oscillations are still present, we also expect a comparable overlap with the orthonormal template. As for the second subleading contribution (eq. \protect\eqref{eq:Fsl_2}, not shown) we find similar behavior, but opposite in sign, resulting in a small but non-negligible cancellation. }

\label{fig:subleading3d}
\end{figure}

Before computing the dot products let us have a crude first look at the shape functions. They appear quite similar to the 
leading order shape, but somewhat surprisingly, as soon as $\delta$ deviates from zero the sub-leading contributions 
change qualitatively. On the left in figure (\ref{fig:subleading3d}) we have plotted eq. \eqref{eq:Fsl_1} for $p=100$ and $\delta=0$, corresponding 
to a rapidly-oscillating function that maximizes in the collinear limit. On the right however, we plotted the same shape function with 
the phase $\delta=3\pi/2$, which suddenly appears to resemble the local shape. Interestingly, the appearance of local type features 
was also noted in the different context of multi-field DBI models (assuming the standard Bunch-Davies vacuum state) \cite{RPetel}.
Based on this observation one would expect the dot product with the 
local shape to increase as a function of $\delta$ (as opposed to what happens for the leading shape function). 
Of course, the collinear feature is still present, but simply swamped by a strong local feature that appears for $\delta=3 \pi/2$. 
In the previous analysis of the leading term we also found that the dot product with the orthogonal template was $\pi$ out of phase as 
compared to the dot product with the local template and that it was comparable in magnitude. We would expect to find something
similar in this sub-leading DBI case. 

Because of the clear presence of local non-Gaussian features for $\delta>0$, which are actually divergent in the $x_2=0$ or $x_3=0$ limit, we have 
to introduce a cutoff when computing dot products, the same cutoff used to normalize the local template. This cutoff is necessary to reflect that we cannot observe infinite wavenumbers in the CMB. It is given by $k_{\rm min}/k_{\rm max}\sim 10^{-3}$, the largest observable physical scale divided by the smallest observable physical scale. So we will require that $x_2, x_3> \epsilon$, with the cutoff $\epsilon$ equal to $\epsilon=10^{-3}$. 

We find that there is substantial overlap with both the local and orthogonal templates. For the dot product with the local template we obtain
\begin{eqnarray}
\nonumber
(F_{sl(1)}+F_{sl(2)})\cdot F_{\rm local} & = &\frac{1}{2p}\left(9+8\log \epsilon \right)\sin \delta  +\frac{1}{p^2\epsilon} \left[ 2\cos(\delta+2p\epsilon)-2(1-\pi p \epsilon)\cos \delta \right.\\
&& +\left.4 p \epsilon \left(\mathrm{ci}(2 p \epsilon)\sin \delta +\mathrm{si}(2 p \epsilon)\cos \delta \right) \right]+ \mathcal{O}(p^{-2},\epsilon).
\end{eqnarray}
Here $\mathrm{ci}=\mathrm{cosintegral}$ and $\mathrm{si}=\mathrm{sinintegral}$. Note that since $\epsilon \sim10^{-3}$ we should be careful 
identifying the second term as $p^{-2}$ suppressed. For small values of $p$ the first term dominates. 
First of all, remember that a factor of $p^2$ was scaled into the non-Gaussian amplitude $f_{\mathrm{NL}}$, explaining the appearance of
the leading $1/p$ suppression. Considering the case $\delta=\pi/2$, $\epsilon=10^{-3}$, we find that 
$(F_{sl(1)}+F_{sl(2)})\cdot F_{\rm local}\sim -31/p$. Properly comparing this to the dot product of the local template with the  
leading order shape function (which requires scaling in one factor of $p$, to match the normalizations of the amplitude), we find that 
that the overlap between the subleading term and the local shape is almost $42$ times better, which is rather surprising. 

\begin{figure}
\begin{centering}
\includegraphics[scale=0.65]{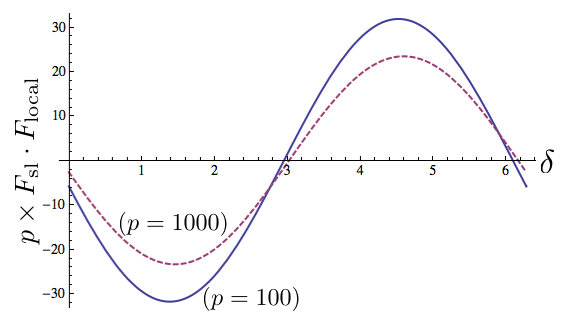}
\par\end{centering}
\caption{The dot product between the subleading ($F_{\mathrm{sl}}=F_{\mathrm{sl(1)}}+F_{\mathrm{sl(2)}}$) and local template (multiplied with $p$) for two values of $p$. Although the dot product is much smaller for $\delta=0$ it is non-zero for both $p$. However, there exist values of the phase for which the dot product (and consequently the leakage or fudge factor) is exactly zero. For such $\delta$ the local template would not be suitable (or at least limited) for detecting any deviation from a BD state (in a DBI model of inflation) and we will have to use the much less capable equilateral template. }
\label{fig:dotproduct_crosssection}
\end{figure}

In figure \ref{fig:dotproduct_crosssection} we have plotted the dot product for two values of $p$ as a function of $\delta$. It clearly shows that 
even though the dot product is smaller for $\delta=0$ it is still relatively large compared to the dot product between the leading term and the 
local template. There exist values for $\delta$ for which the dot product is exactly zero. For those values of the phase the local template 
would not be useful to derive constraints on vacuum state modifications. 

We were unable to obtain an analytical expression for the dot product with the equilateral template and the subleading bispectrum. 
Instead we performed a numerical analysis, which is discussed below.  
Next, we compute the dot product with the equilateral and orthogonal templates. Semi-analytically we obtain from best fitting:
\begin{eqnarray}
(F_{sl(1)}+F_{sl(2)})\cdot F_{\rm ort} & \approx & 5.3 \frac{\sin \delta}{p} + 11.3 \frac{\log p}{p^2}\cos(\delta -0.2 -2.4 \sqrt{p}).
\end{eqnarray}

\begin{figure}
\begin{centering}
\includegraphics[scale=0.65]{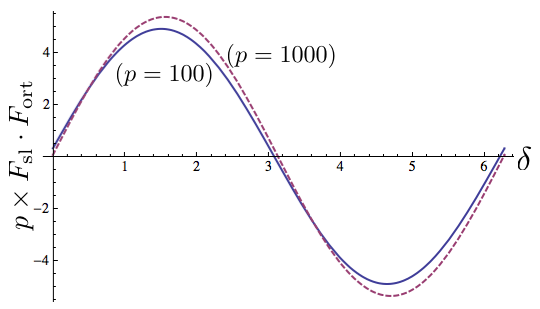}
\par\end{centering}

\caption{The dot product between the subleading and the orthogonal template (multiplied with $p$) for two values of $p$. This dot product is almost completely dominated by the sines and therefore almost zero for $\delta=0$}
\label{fig:dotproduct_crosssection2}
\end{figure}

We compute that $\mathcal{R}(F_{\mathrm{local}},F_{\mathrm{ort}})\sim 1.71$ for $\delta=\pi/2$, indicating that, as expected, the local shape has the best resemblance with the subleading bispectrum.

As in the previous section we use these to obtain the projection factors and then translate these into expressions for the various $f_{\mathrm{NL}}$ 
(which introduces a factor of $p^2$ that was absorbed in the non-Gaussian amplitude): 
\begin{eqnarray}
f_{\mathrm{NL}}^{\rm local} & \simeq & \frac{1}{176.5}\frac{5}{6}\left[ \left(9-32\log 10\right)\sin \delta+\frac{10^3}{p}\left( 2\cos(\delta+ 2\times10^3p)-2(1+ \pi\times10^{-3}p)\cos\delta\right. \right.\nonumber\\
&& \left.+\left. 4\times 10^{-3} p(\mathrm{ci}( 2\times 10^{-3}p)\sin \delta +\mathrm{si}(2\times 10^{-3}p)\cos \delta )\right)\right] \left(\frac{1}{c_{s}^{2}}-1\right)|\beta| p
\label{eq:fnl_sl}
\end{eqnarray}
and
\begin{eqnarray}
f_{\mathrm{NL}}^{\rm ort} & \simeq &\frac{1}{13.8} \frac{5}{3}\left( 5.3 \sin \delta + 11.3 \frac{\log p}{p}\cos(\delta -0.2 -2.4 \sqrt{p})\right)\left(\frac{1}{c_{s}^{2}}-1\right)|\beta| p.
\label{eq:fnl_sl2}
\end{eqnarray}
Once more we distinguish two values of $\delta$.  For $\delta=\pi/2$ and in the small sound speed limit, using that $p\equiv \Lambda_c \, c_s/H$,
the best constraint comes from local non-Gaussinaties and reads
\begin{eqnarray}
|\beta|&\lesssim& \frac{14 H \, c_s}{\Lambda_c}  ,\;\;(\delta=\pi/2),
\label{eq:betac21}
\end{eqnarray} 
which is substantially tighter than the constraint from leading order contributions for the same value of the phase $\delta$ (eq. \eqref{eq:leading_c2}). 
For all the values in between (i.e. $0\leq\delta\leq\pi/2$) the constraints will in general be somewhat weaker. Once the sign of the leakage factor changes, the constraints loose about an order of magnitude, which still exceeds the constraints from the leading term.  
From eq. \eqref{eq:fnl_sl} we find the following constraint for $\delta=0$
\begin{eqnarray}
|\beta|&\lesssim& \frac{72 H \, c_s}{\Lambda_c}, \;\;(\delta=0).
\label{eq:betac22}
\end{eqnarray} 

\begin{figure}
\begin{centering}
\includegraphics[scale=0.65]{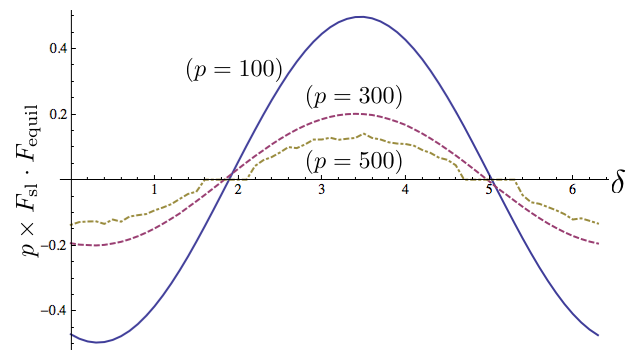}
\par\end{centering}

\caption{Numerical results for the dot product (multiplied by $p$) between the equilateral template and the subleading terms. The dot product gets smaller for larger values of $p$, indicating that the constraints 
using this template are not improved for larger values of $p$ (recall that one $p$ was absorbed into the definition of the associated $f_{\mathrm{NL}}$). In addition, the dot product is already small to begin with, indicating that the equilateral template might
be useful to constrain DBI inflation itself, but is rather poor in constraining DBI inflation with modified initial states. Note that for $p=500$ the numerical computation is at its limit; for $p$ values beyond this the numerical precision is not sufficient to reproduce 
the expected (cosine) oscillations. This holds especially for values that are close to zero. There was no need for a cutoff ($\epsilon$) in these calculations as the dot product was not divergent (as expected) for small values of $x_2$ and $x_3$.}
\label{fig:dp_sl_equil}
\end{figure}

For  $\delta\sim 2.9$ eq. \eqref{eq:betac21} is zero and therefore not useful to place any constraints on $|\beta|$. Similar to the dot products with the leading order terms the dot product with the equilateral template is approximately $\pi/2$ out of phase with the orthogonal and local ones. Obtaining an analytical solution for the dot product  between the equilateral template and the subleading non-Gaussianities turned out to be technically involved, instead we computed these numerically for different values of $p$ and $\delta$. These are shown in figure \ref{fig:dp_sl_equil}. What we find is similar to what we found for the leading order term: it is indeed out of phase with the other two dot products, nor is it enhanced in $p$ (only $\log$-scale). These observations imply that although in principle one could use the equilateral template to constrain $|\beta|$ in the vicinity of those values for which the other two templates would not be able to place a constraint at all, due to the poor overlap with equilateral non-Gaussianities any constraint on $|\beta|$ would be very weak.  It must be noted that the range of values for $\delta$ for which the constraints are weak is limited and it appears, due to small corrections, that the zero points of the the dot products between local and subleading and orthogonal and subleading do not exactly coincide. As a result, one might still be able to obtain a slightly better results using the orthogonal constraints.
Since the equilateral template is only $\log$-scale enhanced by the cutoff, given the constraint on $|\beta|$ from the power spectrum 
(roughly $|\beta|\lesssim10^{-2}$), the signal in the equilateral template will sometimes be dominated by the `regular' DBI contribution, which would allow us to put a constraint on $c_s$ thereby breaking the degeneracy between $c_s$ and $|\beta|$. Unfortunately, this is only true for a small range in $\delta$, effectively only for those values of $\delta$ for which the equilateral contribution from vacuum modification is zero in any case. From best fitting our numerical results, we have been able to deduce that
\begin{eqnarray}
(F_{sl(1)}+F_{sl(2)})\cdot F_{\rm equil} & \sim & -10.8 \cos(\delta - 0.3)\times \frac{\log p}{p^2}.
\end{eqnarray}
Consequently, $f_{\mathrm{NL}}^{\rm equil}$ can be written as 
\begin{eqnarray}
f_{\mathrm{NL}}^{\rm equil} & \simeq & -\frac{5}{3}(0.05+10.8 \cos(\delta - 0.3)\times |\beta| \log p)\left(\frac{1}{c_s^2}-1\right).
\end{eqnarray}

\begin{figure}
\begin{centering}
\includegraphics[scale=0.55]{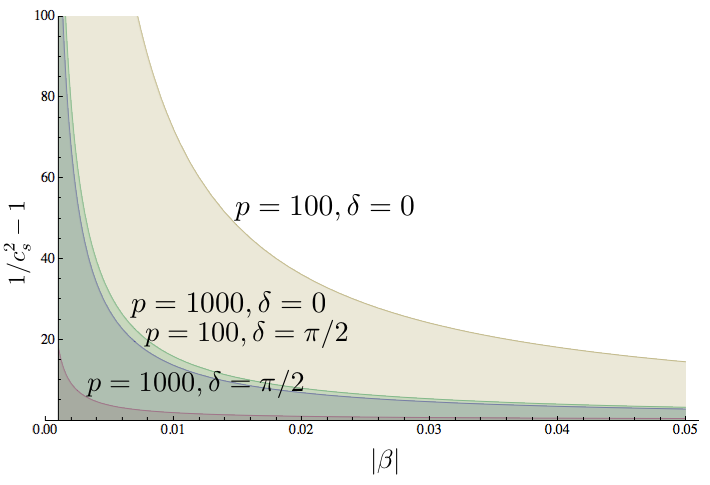}
\par\end{centering}

\caption{Constraints on $|\beta|$ as function of $1/c_s^2-1$ derived from the dot product and the $3\sigma$ levels of local non-Gaussianities for different values of $p$ and $\delta$. These constraints are better than the constraints coming from measurements of the power spectrum.  However, this requires several assumptions due to the appearance of 4 independent variables ($p$, $c_s$, $|\beta|$ and $\delta$). $p$ could be measured independently if one could be able to measure the frequency of the oscillations in the bispectrum. $c_s$ could be determined if one could deduce what type of DBI model initiated inflation. As such, one could break down some of the degeneracy. }
\label{fig:sl_constraints}
\end{figure}

For DBI models of inflation the speed of sound $c_s^{-1}$ can easily be $\mathcal{O}(10)$. In such a scenario for generic values of the phase $\delta$ the constraints become
\begin{eqnarray}
|\beta|&\lesssim&1.4\frac{\Lambda_c}{H}\;\;(\delta=\pi/2), 
\end{eqnarray}
and
\begin{eqnarray}
|\beta|&\lesssim&7.2\frac{\Lambda_c}{H}\;\;(\delta=\pi/2), 
\end{eqnarray}
which implies that for realistic values of the cutoff $|\beta|\lesssim 10^{-3}$. This bests the constraints from the power spectrum by one order of magnitude. As noted however, there exists 
a limited range in $\delta$ for which the constraints are weaker due to the poor overlap with existing templates. Unlike the constraint coming from the bispectrum, the constraints on $|\beta|$ obtained from the power spectrum do not 
depend on the cutoff scale or the phase of the Bogolyubov parameter. As such, the bound on $|\beta|$ for these values of the phase is best constraint by the power spectrum.

\section{Conclusions}

We have analyzed three-point correlators in single-field inflation with a small speed of sound
after introducing a small departure from the standard Bunch-Davies vacuum. Our aim was to
derive the best possible constraints on slightly modified vacua in the specific context of small sound speed models like DBI inflation. 
In previous work \cite{Meerburg} we concluded that the bispectrum is an excellent probe of vacuum state
modifications due to enhancement of the non-Gaussian signal by a factor proportional to (powers of)
the ratio of the cutoff scale and the Hubble parameter $p\equiv \Lambda_c/H c_s$,
which could easily be as large as $10^3$. We have shown that in the small sound speed limit the
powers of $p$ that appear can be as large as $p^3$, which adds one factor of $p$ as compared
to the $c_s=1$ model. In addition the level of non-Gaussianity in small sound speed models is already
large to begin with. Therefore it is not surprising that the constraints that can be derived in such models are in general stronger. 
The computation of the contributions to the local, equilateral and orthogonal template confirmed
this expectation. In general single field models with a small speed of sound the localized
enhancement is proportional to $p^3$, and after projecting onto the local and orthogonal template
one factor of $p$ is lost, ending up with $p^2$ enhancement. The inclusion of an
arbitrary phase in our analysis in the Bogolyubov parameter turns out to be able to reduce the enhancement with
one factor of $p$ for very specific values of the phase. For general single field  models with a small speed of sound
this implies that for generic phases the constraint on the absolute value of the Bogolyubov parameter
equals $|\beta| \lesssim 6\times 10^2 H^2/\Lambda_c^2$, which is actually independent of the speed of sound.
This strong constraint is in fact excluding Bogolyubov parameters linear in $H/\Lambda_c$ in the
small sound speed limit. To be precise, assuming $|\beta| \propto \frac{H^\star}{\Lambda_c}$ one obtains
a ($c_s$ dependent) {\it lower} bound on the ratio $\frac{H}{\Lambda_c}$, which in
the limit of a small sound speed would violate the power spectrum constraint. For the specific value
of the phase $\delta=\pi/2$, the subleading terms are of the same order and have to be included in
the analysis. Similarly, it is the subleading terms that govern the constraint for DBI models, for which
the leading term identically vanishes. 

In the context of DBI models we performed the same analysis, with a reduced enhancement of $p^2$
in general. After projecting onto the different templates and enforcing the observational bounds this resulted
in a constraint on the absolute value of the Bogolyubov parameter
$|\beta| \lesssim 14 H/(\Lambda_c c_s)$, which is linear in $H/\Lambda_c$ and explicitly depends on the speed of sound. This subleading contribution is the one that should also
be taken into account for the specific value of the phase $\delta=\pi/2$ in the general single field case, since this bound is much stronger than the 
bound from the second order terms in the leading expression. For general single field inflation with a small speed of sound this results in a very tight bound on allowed values of $|\beta|$ ranging from $10^{-5}$ to $10^{-3}$. For DBI models of inflation
the constraints are only determined by the subleading term as the leading term vanishes. Again, the inclusion of the phase of the Bogolyubov parameter has the effect that for the specific value of $\delta=0$ the constraint weakens slightly, equal to $|\beta| \lesssim 72 H/\Lambda_c c_s$, but this time remains at the same order in $p$. Overall we conclude that for generic values of $\delta$ and realistic values of the cutoff this still implies 
$|\beta|\lesssim 10^{-3}-10^{-2}$.  For very specific values of the phase, the bounds could vanish altogether. For these special phases one would have to rely on the power spectrum to set a constraint on $|\beta|$.

Our main message is that for single field models with a small speed of sound, including DBI, the constraints
on vacuum state modifications from the bispectrum for generic values of the phase are already better than the power spectrum constraints,
relying on the available templates. The main reason for this is the huge (localized) enhancement that appears after introducing 
a modified initial state and a corresponding cutoff that regularizes the relevant integrals for the bispectrum calculation. 
Improved templates, which would have to be sensitive to the oscillatory
nature of most of the contributions to the bispectrum, are guaranteed to tighten these constraints significantly.
As was previously noted \cite{Meerburg}, the oscillatory nature of the signal in momentum space
suggests that specific, perhaps observable, features could appear in three- and two-dimensional position space.
In any case it would be worthwhile to generalize the range of available
non-Gaussian shapes that can be compared to the data, including oscillatory signals, on which we hope to report in the future.

Our results confirm that higher-derivative corrections
are excellent probes for departures from the standard Bunch-Davies vacuum state.
Throughout this paper we assumed that the combination $|k_1 \eta_0|$ is independent of the actual comoving
momenta involved and equal to $\Lambda_c/H$, in the spirit of the New Physics Hypersurface approach to vacuum state modifications. The reason for this was scale-invariance of the bispectrum, which we relied on to allow for comparison with the available (scale-invariant) template shapes. Fixing $\eta_0$ instead, as one would do in a Boundary Effective Field Theory approach to vacuum state modifications, immediately results in a scale-dependent bispectrum. It would be interesting to study such scale-dependent scenarios and determine to what extent (future) analysis of 3d large scale structure 
or 2d CMB data can constrain bispectrum departures from scale-invariance \cite{NG-scaledep}. 

As reported, the bispectrum or three-point function is extremely sensitive to initial vacuum state modifications in the presence of a higher-derivative operator, and there is no reason to think this could not similarly be true for all higher $n$-point functions. A more general perturbative analysis, including higher $n$-point functions \cite{trispectrum, ChenShiu-etal}, might lead to a hierarchy of (theoretical) constraints on vacuum state modifications, 
perhaps ruling out departures from the standard Bunch-Davies state altogether. Indeed, for the four-point function (trispectrum) the effect of a vacuum 
state modification was recently studied in \cite{ChenShiu-etal}, again showing a strong sensitivity.  We hope that ongoing future work in this direction 
can help us further understand and identify the phenomenological and theoretical constraints the inflationary vacuum state has to satisfy. 

\section{Acknowledgements}
This research was supported in part by a VIDI and a VICI Innovative Research Incentive Grant from the Netherlands Organisation for ScientiÞc Research (NWO).  PDM was supported by the Netherlands Organization for Scientific Research (NWO), NWO-toptalent grant 021.001.040. The research of JPvdS is financially supported by Foundation of Fundamental Research on Matter (FOM) grant 06PR2510. 
\appendix
\section{Full Bispectrum}
In this section we will compute the full bispectrum coming from all terms in the interaction Hamiltonian and for general but $k$-independent $\beta$ with an arbitrary phase, i.e. $\beta=|\beta| \exp{(i\delta)}$. 

We have already computed the leading-order contribution. In the colinear limit $\tilde{k}_j\rightarrow0$ this reduces to 
\begin{eqnarray}
\lim_{\tilde{k}_{j}\rightarrow0}\langle\zeta_{k_{1}}\zeta_{k_{2}}\zeta_{k_{3}}\rangle_{\mathrm{nBD(1)}} & = & \frac{1}{24}(2\pi)^{3}\frac{H^{4}|\beta|}{\epsilon^{2}c_{s}^{4}} \left( c_{s}^{2}-1+\frac{2\lambda c_{s}^{2}}{\Sigma} \right) \delta^{(3)}(\sum\vec{k}_{i})\times\\\nonumber
&&(\eta_0 c_s)^3\left[\frac{1}{k_1k_2(k_1+k_2)}+\frac{1}{k_2k_3(k_2+k_3)}+\frac{1}{k_1k_3(k_1+k_3)}\right]\sin\delta \label{eq_leading_limit}.
\end{eqnarray}

Performing a similar analysis as done for the first term on the remaining terms in the the interaction Hamiltonian yields for the second term:
\begin{eqnarray*}
\langle\zeta_{k_{1}}\zeta_{k_{2}}\zeta_{k_{3}}\rangle_{\mathrm{nBD(2)}} & = &- \frac{1}{16}(2\pi)^{3}\frac{H^{4}|\beta|}{\epsilon^{2}c_{s}^{4}}(\epsilon-3+3c_{s}^{2})\delta(\sum\vec{k}_{i})\times\\
&  & \frac{1}{k_{1}k_{2}k_{3}}\sum_{j}\left[\left(\frac{\tilde{k}_{j}-k_{j}}{k_{j}^{2}}+\frac{\tilde{k}_{j}+k_{j+1}}{k_{j+1}^{2}}+\frac{\tilde{k}_{j}+k_{j+2}}{k_{j+2}^{2}}\right)\left(\frac{\cos \delta-\cos(\tilde{k}_{j}c_{s}\eta_{0}+\delta)}{\tilde{k}_{j}^{2}}\right)\right.\\
&  & +\left.\left(\frac{1}{k_{j}}-\frac{1}{k_{j+1}}-\frac{1}{k_{j+2}}\right)\frac{\eta_{0}c_{s}\sin(\tilde{k}_{j}c_{s}\eta_{0}+\delta)}{\tilde{k}_{j}}\right].
\end{eqnarray*}
In the colinear limit $\tilde{k}_{j}\rightarrow0$ we obtain
\begin{eqnarray*}
\lim_{\tilde{k}_{j}\rightarrow0}\langle\zeta_{k_{1}}\zeta_{k_{2}}\zeta_{k_{3}}\rangle_{\mathrm{nBD(2)}} & = & \frac{1}{32}(2\pi)^{3}\frac{H^{4}|\beta|}{\epsilon^{2}c_{s}^{4}}(\epsilon-3+3c_{s}^{2})\delta(\sum\vec{k}_{i})\\
&&\eta_{0}c_{s}\frac{(k_{1}^2+k_{1}k_{2}+k_{2}^2)\left((k_1k_2(k_1+k_2)\eta_0 c_s\cos \delta -2(k_{1}^2+k_{1}k_{2}+k_{2}^2)\sin \delta \right)}{k_{1}^{3}k_{2}^{3}(k_{1}+k_2)}, \end{eqnarray*}
plus cyclic permutations. There are two things of note here. First of all, it is clear that the subleading term in this limit has two contributions that are out of phase. Secondly, the term proportional to $\sin \delta$ is only linearly enhanced. Therefore, for large values of the cutoff this term is negligible. Consequently, this implies that the leading and subleading are out of phase in their enhancement. Although this observation seems preliminary given that we are considering the colinear limit, it turns out as shown in section 5, that this is true for the full domain. 

The third term yields\begin{eqnarray}
\langle\zeta_{k_{1}}\zeta_{k_{2}}\zeta_{k_{3}}\rangle_{\mathrm{nBD(3)}} & = & \frac{1}{16}(2\pi)^{3}\frac{\epsilon}{c_{s}^{2}}(\epsilon-2s+1-c_{s}^{2})\frac{H^{10}}{\dot{\phi}_{0}^{6}}\frac{1}{c_{s}^{3}}\delta(\sum\vec{k}_{i})\frac{(k_{1}^{2}+k_{2}^{2}+k_{3}^{2})}{k_{1}^{3}k_{2}^{3}k_{3}^{3}}\times\nonumber\\
&  & \mathcal{R}e\Biggl[i\beta\sum_{j}\int_{\eta_{0}}^{0}d\eta\left[\frac{1}{\eta^{2}}-i\tilde{k}_{j}c_{s}\frac{1}{\eta}+(k_{j}k_{j+1}+k_{j}k_{j+2}-k_{j+1}k_{j+2})c_{s}^{2}\right.\nonumber\\
&  & \left.-ik_{1}k_{2}k_{3}c_{s}^{3}\eta\right]\mathrm{exp}( i\tilde{k}_{j}c_{s}\eta ) \Biggl] . \label{bd3}
\end{eqnarray}
This can be integrated to find \begin{eqnarray*}
\langle\zeta_{k_{1}}\zeta_{k_{2}}\zeta_{k_{3}}\rangle_{\mathrm{nBD(3a)}} & = & -\frac{1}{32}(2\pi)^{3}\frac{H^{4}|\beta|}{\epsilon^{2}c_{s}^{4}}(\epsilon-2s+1-c_{s}^{2})\delta(\sum\vec{k}_{i})\frac{1}{k_{1}^{3}k_{2}^{3}k_{3}^{3}}\times\\
&  & \sum_{j}\left(k_1^2+k_{2}^2+k_{3}^2\right)\left[k_{1}k_{2}k_{3}\left(-\frac{\eta_{0}c_{s}\sin(\tilde{k}_{j}c_{s}\eta_{0}+\delta)}{\tilde{k}_{j}}\right)+\right.\\
& &\left.\tilde{k}_{j}\left(\lim_{\eta\rightarrow0}\frac{\sin(\tilde{k_j}c_s\eta+\delta)}{c_s\eta}-\frac{\sin(\tilde{k}_{j}c_{s}\eta_{0}+\delta)}{c_{s}\eta_{0}}\right) +\right.\\
&  & \left. \left( k_1k_2k_3+\tilde{k}_j(k_{j}k_{j+1}+k_{j}k_{j+2}-k_{j+1}k_{j+2}) \right) \frac{\cos \delta -\cos( \tilde{k}_{j}c_{s}\eta_{0}+\delta)}{\tilde{k}_{j}^2}\right].
\end{eqnarray*}
The contribution from the third line would be infinite in the limit $\eta\rightarrow0$ and therefore would produce a divergent result.  However, the limit of the conformal time to zero is not an actual limit since this would suggest we would have infinite time to 
observe all scales. This limit is usually taken for convenience since it turns out that in most cases it does not add power to the 3-point function. In order to constrain this term one should take the limit $\eta\rightarrow \eta^*$ where $\eta^*$ is the conformal time at 
CMB formation. In this way, this terms is finite and negligible compared to all other contributions.  
We can investigate the behavior of this bispectrum in the limit $\tilde{k}_{j}\rightarrow0$
to obtain
\begin{eqnarray*}
\lim_{\tilde{k}_{j}\rightarrow0}\langle\zeta_{k_{1}}\zeta_{k_{2}}\zeta_{k_{3}}\rangle_{\mathrm{nBD(3)}} & = & \frac{1}{32}(2\pi)^{3}\frac{H^{4}|\beta|}{\epsilon^{2}c_{s}^{4}}(\epsilon-2s+1-c_{s}^{2})\delta(\sum\vec{k}_{i})\\
&& \hspace{-0.5in} \eta_{0}c_{s}\frac{(k_{1}^2+k_{1}k_{2}+k_{2}^2)\left((k_1k_2(k_1+k_2)\eta_0 c_s\cos \delta -2(k_{1}^2+k_{1}k_{2}+k_{2}^2)\sin \delta \right)}{k_{1}^{3}k_{2}^{3}(k_{1}+k_2)}, \end{eqnarray*}
plus cyclic permutations, making this limit exactly similar and opposite in sign to the previous collinear limit. In effect it implies that for all values of the phase $\delta$ the collinear limit is zero for DBI models of inflation. 

For the last term in the interaction Hamiltonian we obtain \begin{eqnarray}
\langle\zeta_{k_{1}}\zeta_{k_{2}}\zeta_{k_{3}}\rangle_{\mathrm{nBD(4)}} & = & -\frac{1}{16}(2\pi)^{3}\frac{H^{4}}{\epsilon c_{s}^{4}}\delta(\sum\vec{k}_{i})\frac{|\beta|}{k_{1}^{3}k_{2}^{3}k_{3}^{3}}\times\nonumber\\
&  & \Biggl[\sum_{j}\frac{\cos \delta-\cos(\tilde{k}_{j}c_{s}\eta_{0}+\delta)}{\tilde{k}_{j}}\left(\frac{1}{2}(k_{1}^{4}+k_{2}^{4}+k_{3}^{4})-(k_{1}^{2}k_{2}^{2}+k_{2}^{2}k_{3}^{2}+k_{3}^{2}k_{1}^{2})\right)\nonumber\\
&  & +\sum_{j}\left(-\frac{c_{s}\eta_{0}\sin(\tilde{k}_{j}c_{s}\eta_{0}+\delta)}{\tilde{k}_{j}}+\frac{\cos \delta -\cos(\tilde{k}_{j}c_{s}\eta_{0}+\delta)}{\tilde{k}_{j}^{2}}\right)\times\left(k_{j}^{4}(k_{j+1}+k_{j+2})\right.\nonumber\\
&  & +k_{j+1}^{4}(k_{j+2}-k_{j})+k_{j+2}^{4}(k_{j+1}-k_{j})+k_{j}^{2}k_{j+2}^{2}(k_{j}-2k_{j+1}-k_{j+2})\nonumber\\
&  & \left.+k_{j+1}^{2}k_{j+2}^{2}(2k_{j}-k_{j+1}-k_{j+2})+k_{j}^{2}k_{j+1}^{2}(k_{j}-k_{j+1}-2k_{j+2})\right)\Biggl].\end{eqnarray}

\section{Template normalization and previous results}
The projection between two shapes $F_{X}$ and $F_{Y}$ is determined via
the dot product
\begin{eqnarray}
F_{X}\cdot F_{Y} & \equiv & \int_{\Delta_{k}}dx_{2}dx_{3}F_{X}(x_{2},x_{3})F_{Y}(x_{2},x_{3})x_{2}^{4}x_{3}^{4},
\label{eq:dotproduct_old}
\end{eqnarray}
first introduced in \cite{BCZ2004}. This was previously revised and replaced with the improved dot product (eq. \eqref{eq:correlator}) in \cite{Fergusson}. Although this revised `correlator' improves the 
results compared to the observed 2-dimensional projected dot products, in the case of scale-invariance the differences are marginal. 
The limits of integration are given by the triangle domain, i.e. $\sum\vec{k}_{i}=0$,
which corresponds to $0\leq x_{2}\leq1$ and $1-x_{2}\leq x_{3}\leq1$.
One can then define a cosine between two different shapes to `measure' how similar two shapes with different comoving momenta are
\begin{eqnarray}
\cos(F_{X},F_{Y})&\equiv&\frac{F_{X}\cdot F_{Y}}{(F_{X}\cdot F_{X})^{1/2}(F_{Y}\cdot F_{Y})^{1/2}}.
\end{eqnarray}

In previous work we calculated the normalization factors $F_{\rm local}\cdot F_{\rm local}=176.5$ and
$F_{\rm eq}\cdot F_{\rm eq}=7.9.$ The normalization of the newly proposed orthogonal template $F_{\rm ort}\cdot F_{\rm ort}=13.8$.  Note that the normalization of the local template
is actually infinite if one would integrate over the full triangle domain.  This is because most of the signal comes from scales $k\rightarrow0$
which correspond to perturbations with an infinite physical length
scale. These are not observable in the CMB and therefore one truncates the limit of integration
such that only observable scales are included, which
we take to be $x_{2}>0.001$, as this is the ratio between the smallest and largest observable scales.

\end{document}